\documentclass[journal,onecolumn]{IEEEtran}

\usepackage{amsthm}
\usepackage{comment}
\usepackage{setspace}

\usepackage[dvipsnames,table]{xcolor}

\theoremstyle{definition}
\newtheorem{theorem}{Theorem}[]

\newtheorem{claim}{Claim}[]
\newtheorem{lemma}{Lemma}[]

\usepackage{thm-restate}

\newtheorem{corollary}{Corollary}

\theoremstyle{definition}
\newtheorem{observation}{Observation}

\usepackage[utf8]{inputenc} 
\usepackage[T1]{fontenc}
\usepackage{url}
\usepackage{ifthen}
\usepackage{cite}
\usepackage[cmex10]{amsmath} 

\usepackage[utf8]{inputenc}
\usepackage[T1]{fontenc}
\usepackage{url}
\usepackage{ifthen}
\usepackage{cite}
\usepackage[cmex10]{amsmath}

\usepackage{amsthm}
\usepackage{algorithmic, algorithm}
\usepackage[english]{babel}
\usepackage{enumitem}
\usepackage{amssymb}
\usepackage{xcolor}
\usepackage{graphicx}
\usepackage{tikz}
\usepackage[hidelinks]{hyperref}
\usepackage[font=normalsize]{caption}

\theoremstyle{definition}

\newtheorem{problem}{Problem}

\newtheorem{example}{Example}

\usepackage{mathtools}
\DeclarePairedDelimiter\ceil{\lceil}{\rceil}

\newcommand{\cC}{\mathcal{C}}

\newcommand{\cS}{\mathcal{S}}

\newcommand{\N}{\mathbb{N}}

\newcommand{\bfa}{{\boldsymbol a}}

\newcommand{\bfc}{{\boldsymbol c}}

\newcommand{\bfe}{{\boldsymbol e}}

\newcommand{\bfs}{{\boldsymbol s}}

\newcommand{\bfx}{{\boldsymbol x}}
\newcommand{\bfy}{{\boldsymbol y}}

\newcommand{\bfrh}{{\boldsymbol {\rho}}}
\newcommand{\bfth}{{\boldsymbol {\theta}}}
\newcommand{\bftau}{{\boldsymbol {\tau}}}
\newcommand{\bfchi}{{\boldsymbol {\chi}}}

\newcommand\compsinbase{\Sigma_2^{\text{comp}}\setminus{E_2}}
\newcommand\compbin{\Sigma_2^{\text{comp}}}

\newcommand\ombinnsinbase{\Omega_n^2\setminus{E_2}}
\newcommand\ombinall{\Omega^2}

\newcommand\fth{f_{\bfth}}
\newcommand\coptq{\mathcal{C}_q^{\text{opt}}}

\newcommand\drmldadj{\overline{\text{DR}}_{\text{MLD}}}
\newcommand\Dmldadj{\overline{\mathcal{D}}_{\text{MLD}}}

\newcommand\codebinconbase{\mathcal{BC}_2}
\newcommand\symevencodeconbase{\mathcal{BSC}^{2m}}
\newcommand\symcodeconbase{\mathcal{BSC}^m}
\newcommand\symcodeconbasefour{\mathcal{BSC}^4}
\newcommand\symcode{\mathcal{SC}^m}
\newcommand\codeconbase{\mathcal{BC}^m_q}

\newcommand\optavgprob{\text{VAL}_{\text{avg}}}

\newcommand\optminprob{\text{VAL}_{\text{min}}}
\newcommand\optavg{\text{OPT}_{\text{avg}}}
\newcommand\optmin{\text{OPT}_{\text{min}}}
\newcommand\imn{\mbox{Im}_n}
\newcommand\alldist{\Omega_n^q}
\newcommand\drmld{\text{DR}_{\text{MLD}}}
\newcommand\psucc{\mathbf{P}_{\mbox{succ}}}
\newcommand\fmin{f_{\mbox{\small{min}}}}
\newcommand\favg{f_{\mbox{\small{avg}}}}
\newcommand\Dmld{\mathcal{D}_{\text{MLD}}}
\newcommand\AlphabetSize{q}
\newcommand\Sigmaexp{\Sigma_q^{\textrm{comp}}}

\newcommand{\sfA}{\mathsf{A}}
\newcommand{\sfC}{\mathsf{C}}
\newcommand{\sfG}{\mathsf{G}}
\newcommand{\sfT}{\mathsf{T}}
\newcommand{\sfW}{\mathsf{W}}

\def\CC#1#2{\ensuremath{\left(\kern-.3em\left(\genfrac{}{}{0pt}{}{#1}{#2}\right)\kern-.3em\right)}}
\newcommand\ccnk{\CC{n}{q}}

\begin{document}
\doublespacing
\title{Optimizing the Decoding Probability and Coverage Ratio of Composite DNA} 

\author{%
   \IEEEauthorblockN{\textbf{Tomer Cohen}\IEEEauthorrefmark{1}, and \textbf{Eitan~Yaakobi}\IEEEauthorrefmark{1}}
    \IEEEauthorblockA{\IEEEauthorrefmark{1}%
    Faculty of Computer Science, 
    Technion---Israel Institute of Technology, Israel}
    \texttt{tomer.cohen@campus.technion.ac.il, yaakobi@cs.technion.ac.il}
 }

\maketitle

\begin{abstract}
This paper studies two problems that are motivated by the novel recent approach of \emph{composite DNA} that takes advantage of the DNA synthesis property which generates a huge number of copies for every synthesized strand. Under this paradigm, every \emph{composite symbols} does not store a single nucleotide but a mixture of the four DNA nucleotides. 
The first problem studies the expected number of strand reads in order to decode a composite strand or a group of composite strands.
In the second problem, our goal is study how to carefully choose a fixed number of mixtures of the DNA nucleotides such that the decoding probability by the maximum likelihood decoder is maximized.

\end{abstract}
\section{Introduction}\label{sec:intro}
The main challenge of making DNA storage systems competitive relative to existing storage technologies is the synthesis cost. The simplest and straightforward approach to reduce the cost is to increase the volume of data coded into a given length of oligos, while the information capacity can be measured by bits/symbol or bits/synthesis-cycle. The naive approach, working over $\sfA, \sfC, \sfG, \sfT$ has a theoretical limit of $\log_24 = 2$ bits/symbol, while using error-correction codes can significantly decrease this limit. For example, the information rate in~\cite{Getal13} and~\cite{osti_1619517} is at most $\log_23 \approx 1.58$ since they imposed every two consecutive symbols to be distinct. 
On the other hand, if additional encoding characters are introduced, it is possible to increase the capacity and thus reduce the cost. 

\emph{Composite DNA symbols} were first introduced in~\cite{anavy_DataStorageDNA_2019,augmented_encoding} to leverage the significant information redundancies built into the synthesis and sequencing technologies. 
A composite symbol is a representation of a position in a sequence that does not store just a single nucleotide, but a mixture of the four nucleotides. That is, a composite symbol can be abstracted as a quartet of probabilities $\{p_\sfA, p_\sfC, p_\sfG, p_\sfT\}$, such that $p_b$ describes the fraction of the nucleotide $b \in \{\sfA, \sfC, \sfG, \sfT\}$ present in the mixture, where $0 \le p_b \le 1$ and $p_\sfA+p_\sfC+ p_\sfG+p_\sfT =1$.

For example, $\sfW = (0.5, 0, 0.5, 0)$ represents a composite symbol in which there is a probability of $0.5, 0, 0.5$, and $0$ of seeing $\sfA, \sfC, \sfG$, and $\sfT$, respectively. In the composite DNA oligo $\sfA\sfW\sfT$, 
two types of DNA sequences will exist in the storage container, namely $\sfA\sfA\sfT$ and $\sfA\sfG\sfT$. When synthesizing a composite strand, a cluster of multiple copies of the strand are being generated simultaneously, such that roughly $p_{b}$ of them contain the nucleotide $b$. Thus, in sequencing, one needs to sequence a fraction of the copies from the cluster and then to estimate the probabilities $\{p_\sfA, p_\sfC, p_\sfG, p_\sfT\}$. 
An extension of composite symbol model to the so-called \emph{combinatorial composite DNA} was suggested by Preuss et al.~\cite{PRYA24}. Their idea was to extend the composite idea to work in the shortmer level rather than the base level. In~\cite{ZC22}, Zhang et al. studied error-correcting codes for a model corresponding to DNA composite and a few more models were recently studied in~\cite{ISIT24_1,ISIT24_2,Amit}.

This work studies two important aspects of the DNA composite model.
The first problem is concerned with the sequencing process of DNA in which strands are read randomly. Since every read provides one base of the composite symbol, it is necessary to read many copies of the strands in order to decode the composite symbol, even in the noiseless case.
A related problem to this problem in this paper asks for the probability distribution to decode the composite symbols, given some fixed number of reads~\cite{PGYA24}, while this paper studies the expected number of reads for successful decoding.

In the second problem, we investigate how to carefully choose the mixtures of the probabilities. Namely, assume one wishes to design a DNA storage system with $m$ composite symbols. 
Then, we study how to choose them so the decoding success probability of the Maximum Likelihood Decoder will be maximized. Similarly to this problem the binomial channel was first studied in ~\cite{WESEL}. Another work~\cite{KYW23} studied a similar problem  where the goal is to find the composite mixtures that maximize the composite channel capacity.

The rest of the paper is organized as follows. In Section~\ref{sec:defcomp}, we introduce the composite channel which models composite symbols in DNA strands.
Section~\ref{sec:seq} presents the first problem studied in this paper which asks for the expected number of reads in order to successfully decode the composite symbols. This is studied for a single or multiple strands as well as using an error-correcting code.
Next we study the second problem,
in section~\ref{sec:defs} we define the problem of finding the best composite symbols to maximize the decoding success probability. In Section~\ref{sec:MLD}, we study the the maximum likelihood decoder and its properties. In Section \ref{sec:basic} we focus on finding optimal composite alphabets over general alphabets, for the extreme cases of a small composite alphabet or a very large one and Section~\ref{sec:binary} studies alphabets over the binary case. 
Omitted proofs in the paper appear in the appendix.

\section{The Composite DNA Channel}\label{sec:defcomp}

We model the composite DNA synthesis process as a communication channel, which we refer to as the \emph{composite DNA channel}. In this channel, it is possible to transmit symbols over the $q$-ary alphabet $\Sigma_q\triangleq\{1,\ldots,q\}.$ However, the information that the transmitter aims to convey is \emph{not} a single symbol over $\Sigma_q$, but some \emph{distribution} $\bfrh$ over this alphabet. This information is conveyed by multiple transmissions of symbols over the channel, so the receiver eventually aims to decode the distribution $\bfrh$. This channel is an extension to the binomial channel that was first introduced and studied in \cite{WESEL} and later studied in \cite{KYW23}.

We refer to $\Sigma_q$ as the \emph{base alphabet}, and its extended \emph{composite alphabet}, denoted by $\Sigmaexp$, is defined as the set of all possible  distributions $\bfrh =(\rho_1,\dots,\rho_\AlphabetSize) $ over $\Sigma_q$, i.e., $$\Sigmaexp\triangleq\big\{\bfrh \in \mathbb{R}_{\geq0}^{q}:  \sum_{i=1}^q{\rho_i}=1\big\}.$$ 
For example, if $q=2$, it holds that $\Sigma_2^{\textrm{comp}}=\{(a,1-a) : 0\leq a\leq 1\}$.

In the following section, by considering a special set of distributions from $\Sigmaexp$, we consider the setup where the number of transmissions is \emph{not} fixed, so the user transmits symbols until the decoder can successfully decode the information. Moreover, we consider the case where the user information is not a single composite symbol but a sequence of composite symbols that is sent through parallel basic composite DNA channels for each of the indices. Next in Section~\ref{sec:basic} we try to find optimal alphabets from $\Sigmaexp$ in terms of the decoding probability.

\section{The Coverage Depth Problem for \\Composite DNA}\label{sec:seq}

Given the base alphabet $\Sigma_q$, we define the \emph{$\omega$-composite alphabet}, denoted by $\Sigma_q^{\omega}$, as all distributions from $\Sigmaexp$ with support size ${\omega}$ and identical probability for all the nonzero probabilities (which is $\frac{1}{{\omega}}$). We refer to ${\omega}$ as the \emph{combinatorial factor}. Note that there are exactly $\binom{q}{{\omega}}$ such distributions in $\Sigma_q^{\omega}$ and since the value of every nonzero probability is $\frac{1}{{\omega}}$, it is enough to indicate for every such distribution its support set and thus we define $\Sigma_q^{\omega}$ by $$\Sigma_q^{\omega}\triangleq\left\{A\subseteq\Sigma_q : |A|={\omega}\right\}.$$

\begin{example}\label{ex:comb_factor_symbols}
The set of all symbols of combinatorial factor $2$ over the DNA alphabet contains all the pairs over $\sfA,\sfC,\sfG,\sfT$. There are $\binom{4}{2}$ such pairs and are shown in Fig. \ref{fig:comp_alp_2}.

\begin{figure}[H]
    \centering
    \includegraphics[width=0.5\linewidth]{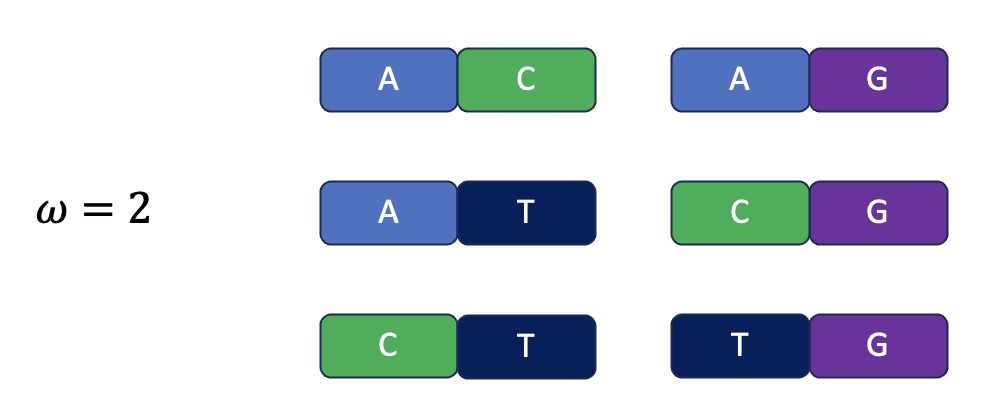}
    \caption{All $2$-composite pairs over $(\sfA,\sfC,\sfG,\sfT)$.}
    \label{fig:comp_alp_2}
\end{figure}
\end{example}

The goal in this section is to transmit a length-$\ell$ sequence of distributions over $\Sigma_q^{\omega}$. A \emph{single transmission} of a sequence $\boldsymbol{s}=(s_1,\dots,s_{\ell})\in(\Sigma_q^{\omega})^{\ell}$ is a sequence  $\mathbf{X}^{\boldsymbol{s}}=(X^{s_1},\dots,X^{s_{\ell}})$ of length $\ell$ over $\Sigma_q$ where $X^{s_i}$ is a random variable of uniform distribution over the set $s_i$. An \emph{$n$-transmission} of $\boldsymbol{s}\in(\Sigma_q^{\omega})^{\ell}$ is a length-$n$ sequence of sequences $\mathbf{X}^{\boldsymbol{s}}_n=(\mathbf{X}^{\boldsymbol{s},1},\dots,\mathbf{X}^{\boldsymbol{s},n})$ of independent single transmissions of $\boldsymbol{s}$.

\begin{example}\label{ex:seq_com_fact}
The sequence $$\boldsymbol{s}=(\{\sfA,\sfC\},\{\sfG,\sfT\},\{\sfC,\sfT\},\{\sfA,\sfG\})$$ is in $(\Sigma_4^2)^{4}$. Each index in the sequence can be visualised as a pair from Fig. \ref{fig:comp_alp_2}. An illustration for the sequence is shown in Fig.~\ref{fig:s_visual}.

\begin{figure}[H]
    \centering
    \includegraphics[width=0.5\linewidth]{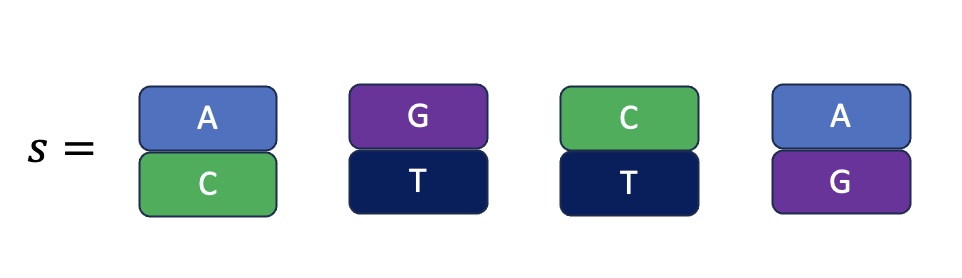}
    \caption{An illustration of the sequence $\boldsymbol{s}$.}
    \label{fig:s_visual}
\end{figure}

 A single transmission of $\boldsymbol{s}$ consists of exactly one symbol for every index of the sequence. For example, the sequence  $(\sfC,\sfG,\sfC,\sfG)$ is a single transmission of $\boldsymbol{s}$. A $5$-transmission for the sequence consists of $5$ single transmissions. It can be visualised using the alphabet in Fig. \ref{fig:comp_alp_2}. In Fig. \ref{fig:trans5_s}, the first row shows the original sequence $\boldsymbol{s}$. Every other row is a different transmission.

\begin{figure}[H]
    \centering
    \includegraphics[width=0.6\linewidth]{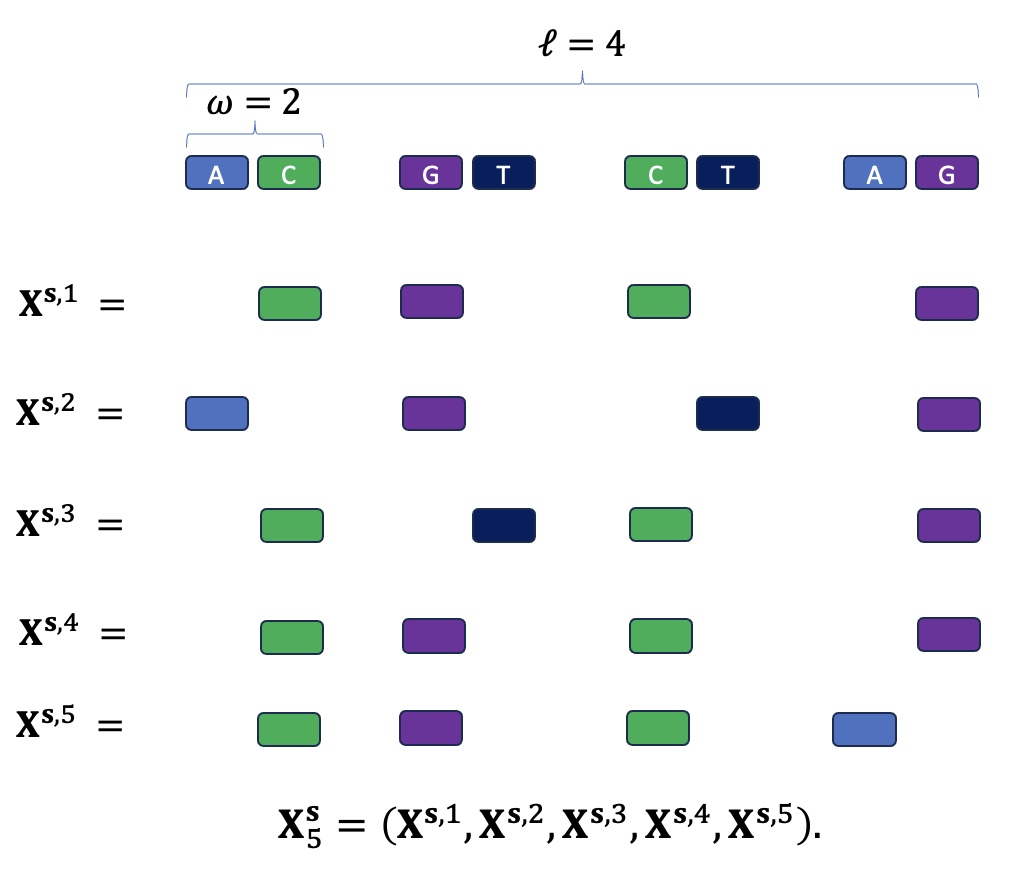}
    \caption{An illustration for a $5$-transmission of $\boldsymbol{s}$.}
    \label{fig:trans5_s}
\end{figure}

\end{example}

The \emph{observed set} of an $n$-transmission $\boldsymbol{\mathbf{X}}^{\boldsymbol{s}}_n$, denoted by $\Gamma(\boldsymbol{\mathbf{X}}^{\boldsymbol{s}}_n)$, is the set of all received single transmissions, i.e.,
$\Gamma(\boldsymbol{\mathbf{X}}^{\boldsymbol{s}}_n)\triangleq\{\mathbf{X}^{\boldsymbol{s},i}\}_{i=1}^{n}\subseteq (\Sigma_q)^\ell.$
Since a single transmission can be transmitted more than once, the size of $\Gamma(\boldsymbol{\mathbf{X}}^{\boldsymbol{s}}_n)$ might be smaller than $n$. The projection of the observed set on the $i$-th index, denoted by $O_i(\Gamma(\boldsymbol{\mathbf{X}}^{\boldsymbol{s}}_n))$, is the set of all symbols that were received for the $i$-th index. An observed set \emph{recovers} the sequence $\boldsymbol{s}=(s_1,\dots,s_{\ell})\in(\Sigma_q^{\omega})^{\ell}$ if the projection on any index $i$ is equal to the set $s_i$, i.e., all ${\omega}$ symbols were observed in each index. Denote by $X_{\boldsymbol{s}}$ a random variable that indicates the number of single transmissions until it is possible to recover the sequence $\boldsymbol{s}$ by the observed set.

\begin{example}
Following Example~\ref{ex:seq_com_fact} for the sequence $\boldsymbol{s}=(\{\sfA,\sfC\},\{\sfG,\sfT\},\{\sfC,\sfT\},\{\sfA,\sfG\})$, consider the same $5$-transmission $\mathbf{X}^{\boldsymbol{s}}_5$ from Example \ref{ex:seq_com_fact}. In Fig. \ref{fig:obs_set_ex}, the column $i$ represents 
$O_i(\Gamma(\boldsymbol{\mathbf{X}}^{\boldsymbol{s}}_n))$, i.e., all the symbols that have been received until the $n$-th transmission.

\begin{figure}[H]
    \centering
    \includegraphics[width=0.6\linewidth]{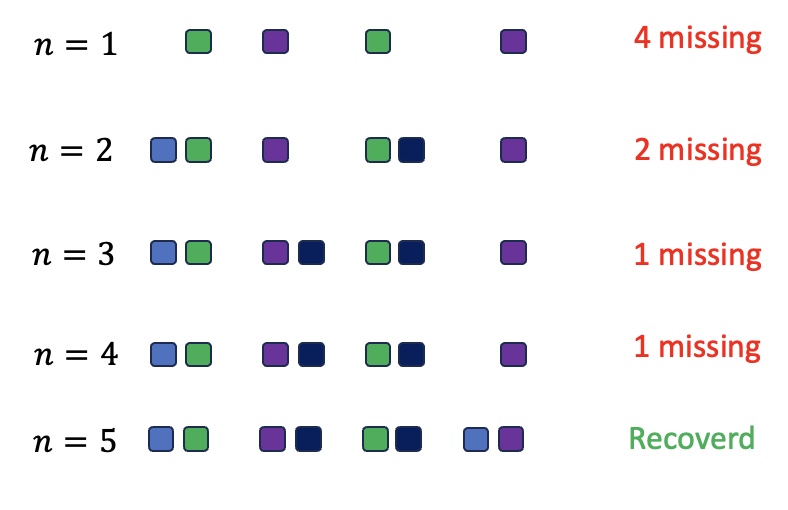}
    \caption{The changing observed set of the $5$-transmission.}
    \label{fig:obs_set_ex}
\end{figure}

The 5-transmission $\mathbf{X}^{\boldsymbol{s}}_5$ recovers $\boldsymbol{s}$ since each index has received all the symbols by the $5$-th transmission.
However, $\mathbf{X}^{\boldsymbol{s}}_4$ does not recover $\boldsymbol{s}$ as the second symbol of the $4$-th index is sent only in the last transmission.
Hence, $X_{\boldsymbol{s}}=5$

\end{example}

From symmetry, we have that for any $\boldsymbol{s,s'}\in(\Sigma_q^{\omega})^{\ell}$ it holds that $X_{\boldsymbol{s}}$ and $X_{\boldsymbol{s'}}$ have the same distribution and therefore, we denote by ${E}(\ell,{\omega},q)$ the expected value of $X_{\boldsymbol{s}}$ for any $\boldsymbol{s}\in(\Sigma_q^{\omega})^{\ell}$, or in words, the expected number of transmissions that are needed in order to recover a sequence of length $\ell$ over $\omega$-composite symbols. 
Given $q,\omega$ and $\ell$ the \emph{coverage depth problem for composite DNA} refers to finding the value of ${E}(\ell,\omega,q)$, which will be studied in this section, as well as two more variations.
One can note that as long as $q \geq \omega$ the size of $q$ is irrelevant for the problem. Therefore, from now on we denote $E(\ell, \omega)$ for $E(\ell, \omega,q)$ with $q\geq 
\omega $.

In the following section we study the expected value of several random variables. Similar random variables were studied in~\cite{PGYA24} whereas the goal was to analyze their success probabilities.

\begin{problem}\label{prob:exp_w_gen}
Given $\ell\in\mathbf{N}$ and $\omega\in\mathbf{N}$, find the value of $E(\ell,\omega)$.
\end{problem}

The following theorem finds the exact solution for Problem~\ref{prob:exp_w_gen} with $\omega=2$.
\begin{theorem}\label{th:w_2}
For any $\ell\in\mathbb{N}$ it holds that
\[E(\ell,2)=2+\sum_{r=1}^{\ell}\binom{\ell}{r}\frac{(-1)^{r+1}}{2^r-1}.\]
\end{theorem}
\begin{proof}
Let $\boldsymbol{s}=(s_1,\dots,s_{\ell})\in(\Sigma_q^2)^{\ell}$. It holds for any $i\in[\ell]$ that $|s_i|=2$. 
In order to recover the sequence $\boldsymbol{s}$, $\Gamma(\boldsymbol{\mathbf{X}}^{\boldsymbol{s}}_n)$ has to receive exactly two symbols in any index. Denote by $X_i$ a random variable of the first transmission in which $|O_i(\Gamma(\boldsymbol{\mathbf{X}}^{\boldsymbol{s}}_n))|=2$. It holds that $${X_{\boldsymbol{s}}}=\max_{i\in[\ell]}{X_i},$$ and therefore, 

$$\mathbf{E}[{X_{\boldsymbol{s}}}]=\mathbf{E}[\max_{i\in[\ell]}{X_i}].$$
For every $i\in[\ell]$ every symbol in the $i$-th index of the transmission can be one of the two symbols that represent the index. One symbol is selected uniformly from $s_i$ of size $2$. In the first transmission we will always receive one symbol from every index. After the first transmission with probability $\frac{1}{2}$ we will receive the symbol that we already have, and with probability $\frac{1}{2}$ we will receive the symbol that we haven't received yet. Therefore we have a random variable that can succeed with probability $\frac{1}{2}$ in every step and it calculates the number of steps that are required to succeed, i.e., $$Y_i\sim{\mbox{Geo}(\frac{1}{2})}$$
with $Y_i=X_i -1 $ since the first transmission always gives the first symbol of every index.
Using the maximum-minimums identity we get that

 $$\mathbf{E}[\max_{i\in[\ell]}{Y_i}]=\sum_{r=1}^{\ell}(-1)^{r+1}\sum_{i_1<\cdots <i_r} \mathbf{E}[\min \{Y_{i_1},\ldots,Y_{i_r}\}].$$ 
 For any $i_1<\cdots<i_r$, the random variable $\min\{Y_{i_1},\dots,Y_{i_r}\}$ calculates the number of trials that are required until at least one of the indices succeed. Since the probability for all the indices to fail is $\frac{1}{2^r}$, the probability for success is $1-\frac{1}{2^r}$. The  random variable has a success probability of $1-\frac{1}{2^r}$ and calculates the number of trials that are required to succeed, therefore, has a geometric distribution, i.e.,
 $$\min\{Y_{i_1},\dots,Y_{i_r}\}\sim{\mbox{Geo}(1-\frac{1}{2^r})}.$$

Using properties of the geometric distribution it holds that,
$$\mathbf{E}[\min\{Y_{i_1},\dots,Y_{i_r}\}]=\frac{2^r}{2^r-1},$$
and therefore,
\begin{align*}
\mathbf{E}[{X_{\boldsymbol{s}}}]=&\mathbf{E}[{Y_{\boldsymbol{s}}}]+1\\
&\sum_{r=1}^{\ell}{\binom{\ell}{r}\frac{2^r}{2^r-1}(-1)^{r+1}}+1\\
=&\sum_{r=1}^{\ell}{\binom{\ell}{r}(-1)^{r+1}}+\sum_{r=1}^{\ell}{\binom{\ell}{r}\frac{1}{2^r-1}(-1)^{r+1}}+1\\
=&2+\sum_{r=1}^{\ell}{\binom{\ell}{r}\frac{(-1)^{r+1}}{2^r-1}}.
\end{align*}

\end{proof}

Next we try to find a solution for a general $\omega$. For every $\omega$ we receive the one new symbol in the first transmission. For $\omega=2$, starting the second transmission, the probability to decode the composite symbol in each index behaves like a geometric random variable because the symbol from the first transmission is a `loss' while the missing second symbol is a `win'. However, for every $\omega>2$ we cannot do it since more than $2$ symbols are needed to recover the index. The next theorem provides a solution to Problem~\ref{prob:exp_w_gen} with a different approach that solves the problem for every $\omega$.

Denote for every $w,m\in\N$, $$\gamma_{w,m}\triangleq\frac{1}{w^m}\sum_{i=1}^{w}{\binom{w}{i}}(-1)^{i+1}(w-i)^m.$$
\begin{theorem}
For any $\ell,w,q\in\N$, it holds that 
$$E(\ell,w)=\sum_{m=0}^{\infty}(1-(1-{\gamma_{w,m}})^{\ell}).$$
\end{theorem}
\begin{proof}
Let $\boldsymbol{s}=(s_1,\dots,s_{\ell})\in(\Sigma_q^w)^{\ell}$. It holds for any $i\in[\ell]$ that $|s_i|=w$.
In order to recover the sequence $\boldsymbol{s}$, $\Gamma(\boldsymbol{\mathbf{X}}^{\boldsymbol{s}}_n)$ has to receive all the $w$ symbols in every index.

Denote by $X_i$ a random variable of the first transmission in which $|O_i(\Gamma(\boldsymbol{\mathbf{X}}^{\boldsymbol{s}}_n))|=w$. It holds that
$${X_{\boldsymbol{s}}}=\max_{i\in[\ell]}{X_i}$$ and therefore, $$\mathbf{E}[{X_{\boldsymbol{s}}}]=\mathbf{E}[\max_{i\in[\ell]}{X_i}].$$
For every $i\in[\ell]$ every symbol in the $i$-th index of the transmission is selected uniformly from $s_i$ of size $w$.
Using the maximum-minimums identity get that
$$\max_{i\in[\ell]}{X_i}=\sum_{r=1}^{\ell}(-1)^{r+1}\sum_{i_1<\cdots <i_r} \min \{X_{i_1},\ldots,X_{i_r}\}.$$
For $1\leq r\leq \ell$ and for all $r$ indices $1\leq i_1<\cdots <i_r\leq \ell$, the random variable $\min \{X_{i_1},\ldots,X_{i_r}\}$ is distributed the same from symmetry. Since the random variables $X_1,\dots,X_{\ell}$ are independent it holds that for every $m\in\N$ $$
\mathbf{P}[\min \{X_1,\ldots,X_r\}>m]=\prod_{i=1}^{r}{\mathbf{P}[X_i>m]}.$$ 
The probability $\mathbf{P}[X_i>m]$ is the probability to uniformly sample a word from an alphabet of size $w$ such that it does not contain all its $w$ symbols. 
By the inclusion exclusion principle we get that for every $1\leq j\leq \ell$ it holds that
$$\mathbf{P}[X_j>m]=\frac{1}{w^m}\sum_{i=1}^{w}{\binom{w}{i}}(-1)^{i+1}(w-i)^m = \gamma_{w,m}.$$
As a result for every $m\in\N$ it holds that
$$\mathbf{P}[\min \{X_1,\ldots,X_r\}>m]=\prod_{i=1}^{r}{\mathbf{P}[X_i>m]}=\gamma_{w,m}^r.$$
Using the maximum-minimums identity we get

\begin{align*}
    &\mathbf{P}[\max_{i\in[\ell]}{X_i}>m]=\\
    &=\sum_{r=1}^{\ell}(-1)^{r+1}\sum_{i_1<\cdots <i_r} \mathbf{P}[\min \{X_1,\ldots,X_r\}>m]\\
    &=\sum_{r=1}^{\ell}(-1)^{r+1}\sum_{i_1<\cdots <i_r} \gamma_{w,m}^r=\sum_{r=1}^{\ell}(-1)^{r+1}\binom{\ell}{r} \gamma_{w,m}^r
\end{align*}

Therefore, we get that
\begin{align*}
   \mathbf{P}[\max_{i\in[\ell]}{X_i}>m]=&\sum_{r=1}^{\ell}(-1)^{r+1}\binom{\ell}{r} \gamma_{w,m}^r\\=&1-(1-{\gamma_{w,m}})^{\ell}. 
\end{align*}

Lastly, using the tail sum formula for expectation we get that
\begin{align*}
    \mathbf{E}[\max_{i\in[\ell]}{X_i}]=&\sum_{m=0}^{\infty}{\mathbf{P}[\max_{i\in[\ell]}{X_i}>m]}\\
    =&\sum_{m=0}^{\infty}{\mathbf{P}[\max_{i\in[\ell]}{X_i}>m]}\\
    =&\sum_{m=0}^{\infty}(1-(1-{\gamma_{w,m}})^{\ell}).
\end{align*}

\end{proof}

For $\omega=2$ we have that $$\gamma_{2,m}=\frac{2}{2^m}=\frac{1}{2^{m-1}},$$
and therefore,
$$E(\ell,2)=1+\sum_{m=1}^{\infty}\left(1-(1-\frac{1}{2^{m-1}})^{\ell}\right),$$
which provides another expression for the solution given in Theorem~\ref{th:w_2}. Next we try to find an estimation for the solution of Problem~\ref{prob:exp_w_gen} with $\omega=2$.

\begin{figure}[H]
    \centering
    \fbox{\includegraphics[width=0.75\linewidth]{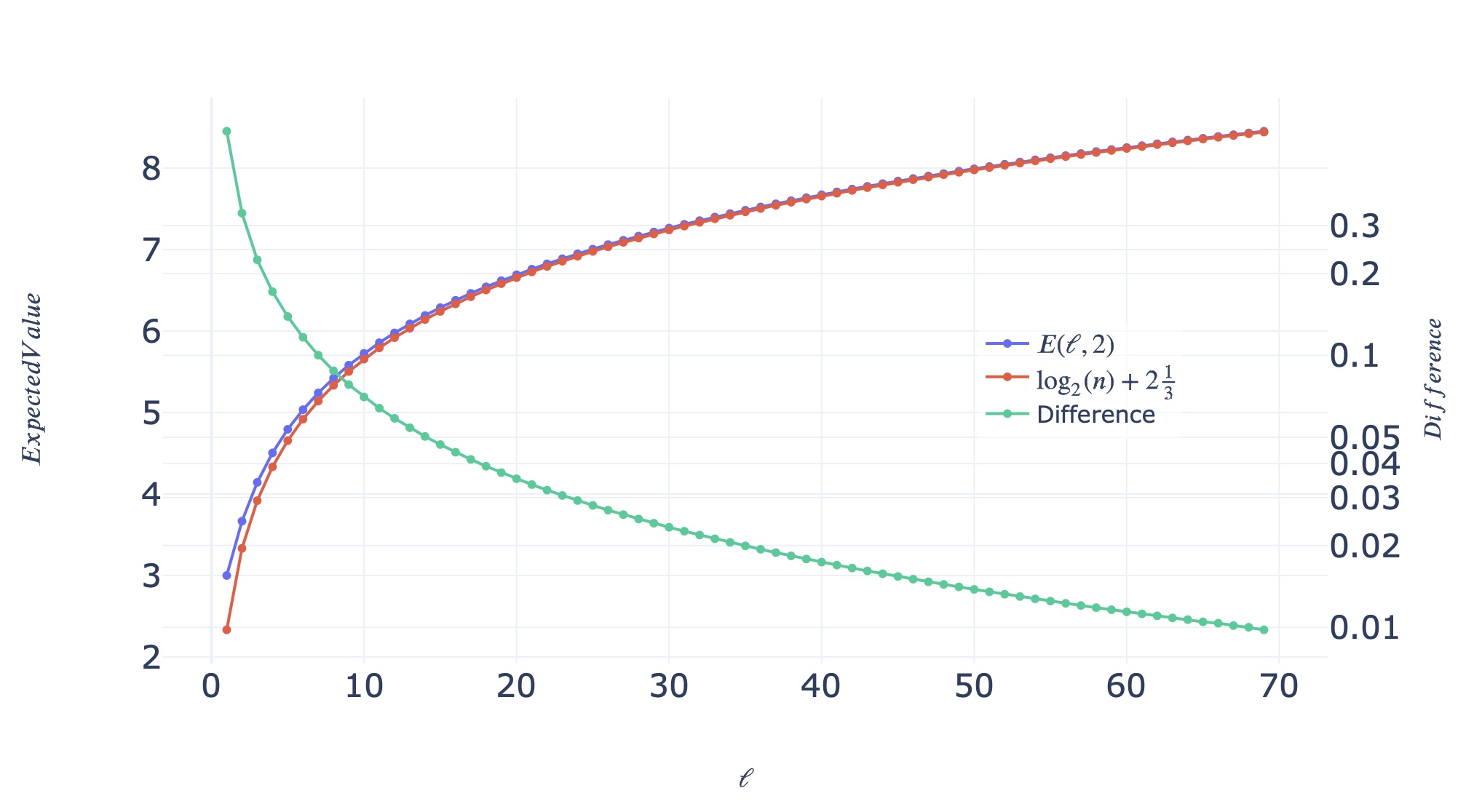}}
    \caption{The value of $E(\ell,2)$ (blue) and $\log_2(\ell)+2\frac{1}{3}$ (red) as a function of $\ell$. The difference in green on logarithmic scale.}
    \label{fig:E_2_graph}
\end{figure}

In Fig. \ref{fig:E_2_graph} we have the value of $E(\ell,2)$ as a function of $\ell$ (in blue). It is important to note that the value of $E(\ell,2)$ is almost exactly $\log_2(\ell)+2\frac{1}{3}$ (in red).

The following lemma is based on analysis in~\cite{RVE} and is crucial for the future estimation.

\begin{lemma}\label{lm:geom_estim}
Let $n\in\mathbf{N}$ and $0\leq p\leq1$.
Let $X_1,X_2,X_3,\dots,X_n$ be IID geometric random variables with success probability of $p$. Denote $$M_n=\max\{X_1,X_2,X_3,\dots,X_n\}.$$ It holds that
\[
\frac{1}{\lambda}\sum_{i=1}^{n}{\frac{1}{i}}<\mathbf{E}[M_n]<1+\frac{1}{\lambda}\sum_{i=1}^{n}{\frac{1}{i}}
\]
for $\lambda=\ln{\frac{1}{1-p}}$.
\end{lemma}
\begin{proof}
    It holds for every $1\leq i\leq n$ that
    \[
    \mathbf{P}[X_i\leq k]=1-(1-p)^k.
    \]
    Hence,
    \[
    \mathbf{P}[M_n\leq k]=\Pi_{i=1}^{n}{\mathbf{P}[X_i\leq k]}=(1-(1-p)^k)^n.
    \]
    Using the tail-sum formula we have that
    \[
    \mathbf{E}(M_n)=\sum_{k=0}^{\infty}{\mathbf{P}[M_n> k]}=\sum_{k=0}^{\infty}{1-(1-(1-p)^k)^n}.
    \]
    Let $\lambda=\ln{\frac{1}{1-p}}$, i.e, $1-p=e^{-\lambda}$. The function $f(x)=1-(1-e^{-\lambda x})^n$ is increasing in $[0,\infty)$. Therefore, it holds that
    \[
    \int_0^{\infty}{f(x)dx}\leq \sum_{i=0}^{\infty}{f(i)}\leq f(0) +\int_0^{\infty}{f(x)dx}.
    \]
    Hence,
\begin{align*}
    \int_0^{\infty}{1-(1-e^{-\lambda x})^n dx} &\leq \sum_{k=0}^{\infty}{1-(1-(1-p)^k)^n} \\&\leq 
    1+\int_0^{\infty}{1-(1-e^{-\lambda x})^n dx}.
\end{align*}
It holds that
\[
\int_0^{\infty}{1-(1-e^{-\lambda x})^n dx}=\frac{1}{\lambda}\sum_{i=1}^{n}{\frac{1}{i}}.
\]
Therefore,
\[
\frac{1}{\lambda}\sum_{i=1}^{n}{\frac{1}{i}}<\mathbf{E}[M_n]<1+\frac{1}{\lambda}\sum_{i=1}^{n}{\frac{1}{i}}.
\]
\end{proof}

The following theorem provides the bounds for Problem~\ref{prob:exp_w_gen} with $\omega=2$.

\begin{theorem}\label{th:iid_geom_approx}
For every $\ell>0$ it holds that
\begin{align*}
    1+\log_2(\ell)+\frac{1}{\ell\ln(2)} & \leq E(\ell,2) \leq 2+ \log_2(\ell)+\frac{1}{\ln(2)},   
\end{align*}
and $2+ \log_2(\ell)+\frac{1}{\ln(2)}\approx\log_2{\ell}+3.443$.
\end{theorem}
\begin{proof}
Using the random variables from Theorem~\ref{lm:geom_estim} it holds that $\{Y_i\}_{i=1}^{\ell}$ that are defined as $Y_i=X_i-1$ are IID geometric random variables with success probability $\frac{1}{2}$. One can confirm that since in the first transmission we always receive the first symbol of all the indices, then in each transmission we can receive the second symbol and recover the index with probability $\frac{1}{2}$.
Using Lemma \ref{th:w_2} we have that
\[
\frac{1}{\ln{2}}\sum_{i=1}^{\ell}{\frac{1}{i}}<E[\max\{Y_1,Y_2,\dots,Y_\ell\}]<1+\frac{1}{\ln{2}}\sum_{i=1}^{\ell}{\frac{1}{i}}.
\]
Since $Y_i=X_i-1$ it holds that
\[
1+\frac{1}{\ln{2}}\sum_{i=1}^{\ell}{\frac{1}{i}}<E[\max\{X_1,X_2,\dots,X_\ell\}]<2+\frac{1}{\ln{2}}\sum_{i=1}^{\ell}{\frac{1}{i}}.
\]
Therefore,
\[
1+\frac{1}{\ln{2}}\sum_{i=1}^{\ell}{\frac{1}{i}}<E(\ell,2)<2+\frac{1}{\ln{2}}\sum_{i=1}^{\ell}{\frac{1}{i}}.
\]
It is known that
\[
\ln(\ell)+\frac{1}{\ell}\leq \sum_{i=1}^{\ell}{\frac{1}{i}}\leq \ln(\ell)+1.
\]
Therefore, it holds that 
\[
\log_2(\ell)+1+\frac{1}{\ell\ln(2)}\leq E(\ell,2) \leq 2+ \log_2(\ell)+\frac{1}{\ln(2)}\approx\log_2{\ell}+3.443.
\]
\end{proof}

Next, we try to find an estimation for the solution of Problem~\ref{prob:exp_w_gen} for general $\omega$. The upper bound is given in the following theorem.

\begin{theorem}
For every $\ell>0$ and $\omega\geq2$ it holds that
\[
E(\ell,\omega)\leq 1+\frac{\log_2{\omega\ell}}{\log_2\left(\frac{\omega}{\omega-1}\right)}+\omega.
\]
\end{theorem}
\begin{proof}
Let $\ell>0$ and $\omega\geq 2$.
It holds that
$$\gamma_{w,m}\triangleq\frac{1}{w^m}\sum_{i=1}^{w}{\binom{w}{i}}(-1)^{i+1}(w-i)^m.$$
One can note that the term is derived using the exclusion inclusion principle. By using union bound it holds that 
    $$\gamma_{w,m}\leq \omega\left(\frac{\omega-1}{\omega}\right)^m.$$
For any $\ell,w,q\in\N$, it holds that 
$$E(\ell,w)=\sum_{m=0}^{\infty}(1-(1-{\gamma_{w,m}})^{\ell}).$$
For every $\mu\in\N$ it holds that
\[
\sum_{m=0}^{\infty}(1-(1-{\gamma_{w,m}})^{\ell})\leq \mu +\sum_{m=\mu}^{\infty}(1-(1-{\gamma_{w,m}})^{\ell})
\]
since all the values are bounded by 1.
Bounding the probability results
\[
(1-(1-{\gamma_{w,m}})^{\ell})\leq \ell \gamma_{w,m} \leq \ell\omega\left(\frac{\omega-1}{\omega}\right)^m
\]
Let $m^*$ be the smallest value of $m$ such that
\[
\ell\omega\left(\frac{\omega-1}{\omega}\right)^m\leq 1.
\]
Note that $m^*$ exists since $\ell\omega\left(\frac{\omega-1}{\omega}\right)^m$ is decreasing as a function of $m$.
In order to find $m^*$ one can note that 
\[
\ell\omega\left(\frac{\omega-1}{\omega}\right)^m\leq 1
\]
if and only if
\[
\left(\frac{\omega-1}{\omega}\right)^m\leq \frac{1}{\omega\ell}
\]
if and only if
\[
m\geq \frac{\log_2{\frac{1}{\omega\ell}}}{\log_2\left(\frac{\omega-1}{\omega}\right)}.
\]
One can observe that
\[
m^*= \ceil*{\frac{\log_2{\frac{1}{\omega\ell}}}{\log_2\left(\frac{\omega-1}{\omega}\right)}}\leq \frac{\log_2{\frac{1}{\omega\ell}}}{\log_2\left(\frac{\omega-1}{\omega}\right)}+1.
\]
Furthermore, by definition,
\[
{\ell\omega\left(\frac{\omega-1}{\omega}\right)^{m^*}}\leq 1.
\]
and therefore,
\begin{align*}
    \sum_{m=0}^{\infty}(1-(1-{\gamma_{w,m}})^{\ell})&\leq m^* +\sum_{m=m^*}^{\infty}(1-(1-{\gamma_{w,m}})^{\ell})\leq m^* +\sum_{m=m^*}^{\infty}{\ell\omega\left(\frac{\omega-1}{\omega}\right)^m}\leq m^* +\frac{{\ell\omega\left(\frac{\omega-1}{\omega}\right)^{m^*}}}{1-\frac{\omega-1}{\omega}}\\
    &\leq m^*+\frac{1}{1-\frac{\omega-1}{\omega}}\leq 1+\frac{\log_2{\frac{1}{\omega\ell}}}{\log_2\left(\frac{\omega-1}{\omega}\right)}+\omega\leq 1+\frac{\log_2{\omega\ell}}{\log_2\left(\frac{\omega}{\omega-1}\right)}+\omega.
\end{align*}
\end{proof}

The following theorem provides a lower bound for Problem~\ref{prob:exp_w_gen} for general $\omega$.
\begin{theorem}\label{th:iid_geom_approx}
For every $\ell>0$ and $\omega\geq2$ it holds that
\[
E(\ell,\omega)\geq\frac{\log_2{\ell}}{\log_2{\frac{\omega}{\omega-1}}}+\frac{\log_2{e}}{\ell \log_2(\frac{\omega}{\omega-1})}.
\]
\end{theorem}
\begin{proof}
Let $\ell>0$ and $\omega\geq 2$. It holds that
$$\gamma_{w,m}=\frac{1}{w^m}\sum_{i=1}^{w}{\binom{w}{i}}(-1)^{i+1}(w-i)^m.$$
One can note that the term is derived using the exclusion inclusion principle. The intersection of all the sets is a lower bound, therefore, it holds that
    $$\gamma_{w,m}\geq \left(\frac{\omega-1}{\omega}\right)^m.$$
    Therefore it holds that
    \begin{align*}
        E(\ell,w)=&\sum_{m=0}^{\infty}(1-(1-{\gamma_{w,m}})^{\ell})\geq\sum_{m=0}^{\infty}(1-(1-\left(\frac{\omega-1}{\omega}\right)^m)^{\ell})
    \end{align*}
    The second term is exactly the expected value for $\max\{X_1,X_2,\dots,X_\ell\}$ with IID $X_i\sim \mbox{Gem}(\frac{1}{\omega})$. Using Lemma~\ref{lm:geom_estim} it holds that
    \begin{align*}
        \sum_{m=0}^{\infty}(1-(1-\left(\frac{\omega-1}{\omega}\right)^m)^{\ell})
        \geq&\frac{1}{\lambda}{\sum_{i=1}^\ell{\frac{1}{i}}}
        \geq\log_{\frac{\omega}{\omega-1}}(\ell)+\frac{1}{\ell \ln(\frac{\omega}{\omega-1})}
        \geq\log_{\frac{\omega}{\omega-1}}(\ell)+\frac{1}{\ell \ln(\frac{\omega}{\omega-1})}\\
        =&\frac{\log_2{\ell}}{\log_2{\frac{\omega}{\omega-1}}}+\frac{\log_2{e}}{\ell \log_2(\frac{\omega}{\omega-1})}.
    \end{align*} 
\end{proof}

\begin{figure}[H]
    \centering
    \fbox{\includegraphics[width=0.5\linewidth]{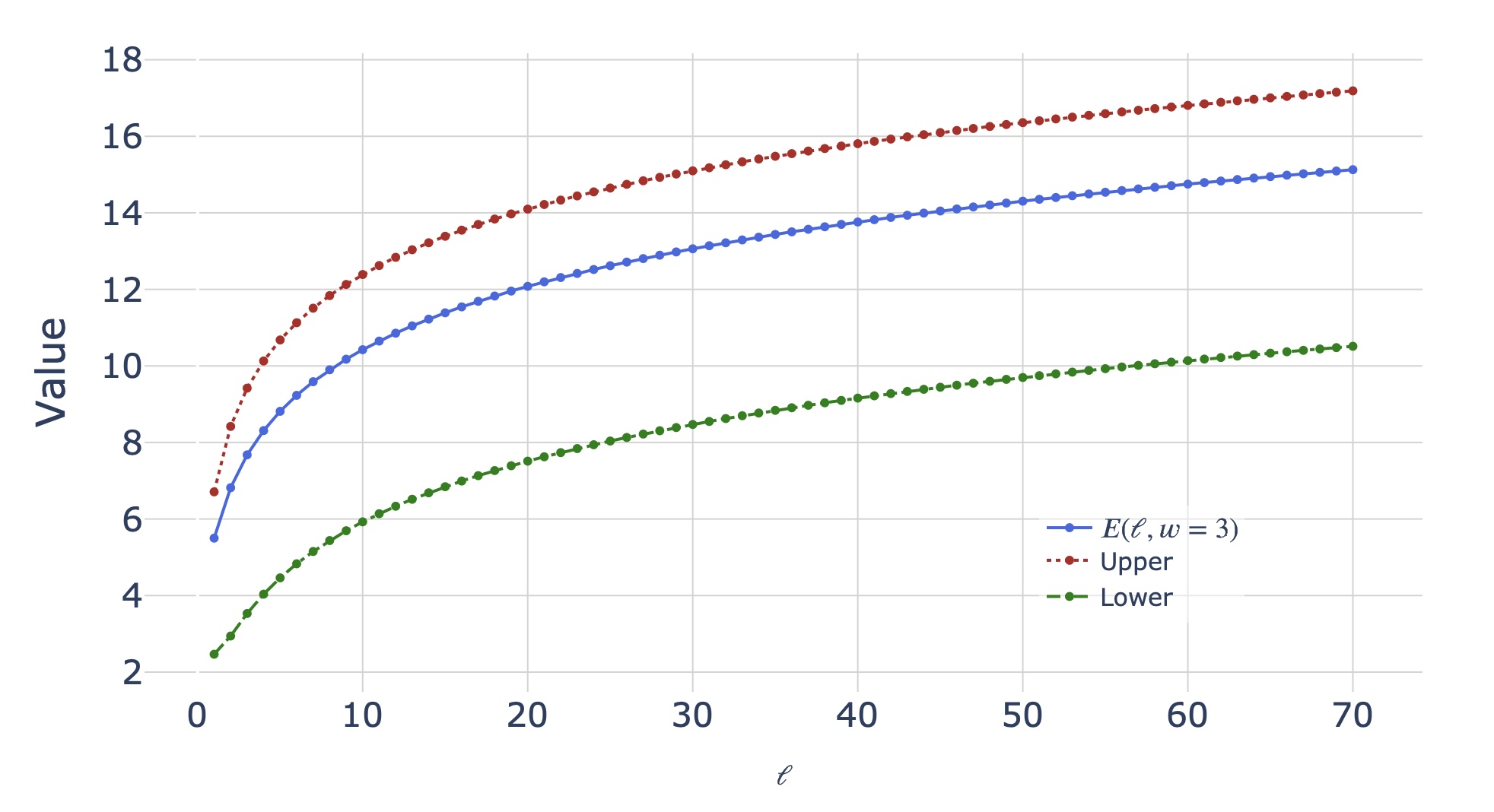}}
    \caption{The value of $E(\ell,3)$ (blue), the lower bound (green) and the upper bound (red).}
    \label{fig:E_3_graph}
\end{figure}

\begin{figure}[H]
    \centering
    \fbox{\includegraphics[width=0.5\linewidth]{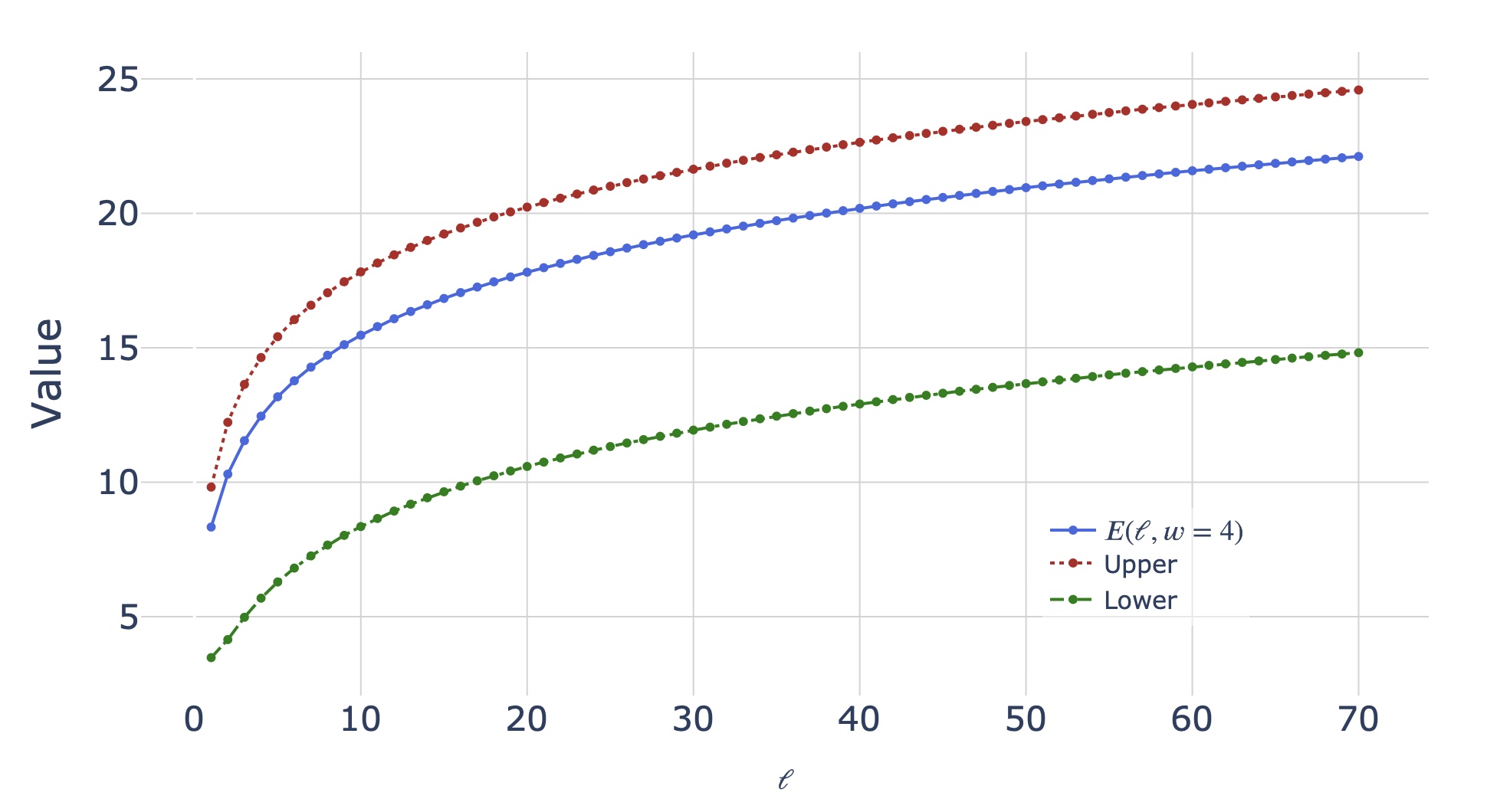}}
    \caption{The value of $E(\ell,4)$ (blue), the lower bound(green) and the upper bound(red).}
    \label{fig:E_4_graph}
\end{figure}

In 
Fig. \ref{fig:E_3_graph} and Fig. \ref{fig:E_4_graph} we have the value of $E(\ell,\omega)$ and the lower and upper bounds from the previous theorems. One can note that the slope is almost identical in all of the graphs but the exact constant can be found empirically.


\begin{corollary}
    It holds that
    \[
    E(\ell,\omega) = a_\omega \log_2(\ell) + C_\omega,
    \]
    with
    \[
    a_\omega=\frac{1}{\log_2{\frac{\omega}{\omega-1}}}.
    \]
\end{corollary}
Next, we consider another setup of this problem. Here we assume that an error-correcting code is applied so it is not necessary to recover all the set entries of $\boldsymbol{s}$, but only a subset of them. Let $\boldsymbol{s}=(s_1,\dots,s_{\ell})\in(\Sigma_q^w)^{\ell}$ and $1\leq r\leq\ell$. Denote by $X_{\boldsymbol{s},r}$ a random variable that indicates the number of single transmissions until it is possible to recover at least $r$ out of the $\ell$ indices in the sequence by the observed set. One can observe that $X_{\boldsymbol{s},\ell}=X_{\boldsymbol{s}}$.
From symmetry we have that for any $\boldsymbol{s,s'}\in(\Sigma_q^w)^{\ell}$, $X_{\boldsymbol{s}},r$ and $X_{\boldsymbol{s'},r}$ have the same distribution and therefore, we denote by ${E}(\ell,w,q;r)$ the expected value of $X_{\boldsymbol{s},r}$ for any $\boldsymbol{s}\in(\Sigma_q^w)^{\ell}$. Similarly, $q$ has no importance in the problem as long as $q\geq \omega$, so we can denote 
$${E}(\ell,\omega;r)=E(\ell,\omega,q;r).$$

\begin{problem}\label{prob:at_least_r}
Given $\ell,\omega,r\in\mathbf{N}$, find the value of $E(\ell,\omega,r)$.
\end{problem}

The following theorem calculates the exact solution for Problem~\ref{prob:at_least_r}.

\begin{theorem}
For any $\ell, \omega, r\in \N$
where
$1\leq r \leq\ell$
it holds that
\begin{align*}
&E(\ell,\omega;r)=\\=&\sum_{j=1}^{\binom{\ell}{r}}\sum_{m=1}^{\ell}\sum_{i=1}^{m}(-1)^{i+j}\binom{\ell}{m}\binom{m}{i}\binom{\binom{m-i}{r}}{j}E(m,w,q).
\end{align*}
\end{theorem}

\begin{proof}
Denote by $X_i$ a random variable of the first transmission in which $|O_i(\Gamma(\boldsymbol{\mathbf{X}}^{\boldsymbol{s}}_n))|=w$.
Let $\ell\in\N$ and $r\leq\ell$, denote by $\mathcal{I}_r^{\ell}$ the set of all $\binom{\ell}{r}$ sorted $r$-vectors of indices from $[\ell]$, i.e., $
 \mathcal{I}_r^{\ell}=\{(i_1,\dots,i_r):1\leq i_1<\dots<i_r\leq\ell\}.$
 Furthermore, we use the lexicographic order on the set $\mathcal{I}_r^{\ell}$. For every $I\in\mathcal{I}_r^{\ell}$ we denote $X_I=\max_{i\in I}{X_i}$.
It holds that
${X_{\boldsymbol{s}, r}}=\min_{I\in\mathcal{I}_r^{\ell}}{X_I}$ and therefore, $\mathbf{E}[{X_{\boldsymbol{s},r}}]=\mathbf{E}[\min_{I\in\mathcal{I}_r^{\ell}}{X_I}]$.
Using the maximum-minimums identity get that
$$ \min_{I\in\mathcal{I}_r^{\ell}}{X_I}=\sum_{j=1}^{\binom{\ell}{r}}(-1)^{j+1}\sum_{I_1<\cdots <I_j\in\mathcal{I}_r^{\ell}} \max \{X_{I_1},\ldots,X_{I_j}\}.$$

It holds that for any $I_1<\cdots <I_r\in\mathcal{I}_r^{\ell}$
\[
\max \{X_{I_1},\ldots,X_{I_r}\}=\max_{i\in \bigcup_{b=1}^{j}{I_b}}{X_b}
\]
We denote by $\alpha(\ell,r,m,j)$ the number of sets of size $j$ of $r$-arrays over $\ell$ such that their union is equal and of size $m$, i.e., for any $m$ indices $1\leq i_1<\cdots< i_{m}\leq\ell$ the number of sets $\{{I_1,\dots,I_j}:I_b\in\mathcal{I}_r^{\ell}\}$ such that
\[
\bigcup_{b=1}^{j}{I_b}=\{i_1,\dots,i_{m}\}.
\]
Using inclusion exclusion principle we get that for every $\ell,r,m,j\in\N$ it holds that
\[
\alpha(\ell,r,m,j)=\sum_{i=1}^{m}{\binom{m}{i}}(-1)^{i+1}\binom{\binom{m-i}{r}}{j}.
\]
We can conclude that for every $1\leq j\leq\binom{\ell}{r}$ it holds that
\begin{align*}
&\sum_{I_1<\cdots <I_j\in\mathcal{I}_r^{\ell}} \mathbf{E}[\max \{X_{I_1},\ldots,X_{I_j}\}]=\sum_{m=1}^{\ell}\binom{\ell}{m}\alpha(\ell,r,m,j)\mathbf{E}[\max \{X_1,\ldots,X_r\}]=\sum_{m=1}^{\ell}\binom{\ell}{m}\alpha(\ell,r,m,j)E(m,w,q).
\end{align*}
Therefore,
\begin{small}
\begin{align*}
\mathbf{E}[{X_{\boldsymbol{s},r}}]=&\sum_{j=1}^{\binom{\ell}{r}}(-1)^{j+1}\sum_{I_1<\cdots <I_j\in\mathcal{I}_r^{\ell}} \mathbf{E}[\max \{X_{I_1},\ldots,X_{I_j}\}]
   =\sum_{j=1}^{\binom{\ell}{r}}(-1)^{j+1}\sum_{m=1}^{\ell}\binom{\ell}{m}\alpha(\ell,r,m,j)E(m,w)\\
   =&\sum_{j=1}^{\binom{\ell}{r}}(-1)^{j+1}\sum_{m=1}^{\ell}\binom{\ell}{m}\sum_{i=1}^{m}{\binom{m}{i}}(-1)^{i+1}\binom{\binom{m-i}{r}}{j}E(m,w)
   =\sum_{j=1}^{\binom{\ell}{r}}\sum_{m=1}^{\ell}\sum_{i=1}^{m}(-1)^{i+j}\binom{\ell}{m}\binom{m}{i}\binom{\binom{m-i}{r}}{j}E(m,w).
\end{align*}
\end{small}
\end{proof}

\begin{observation}
    Since we can choose the $\ell-r$ symbols that are not required to be decoded it holds that
    \[
    E(\ell,\omega;r)\leq E(r,\omega).
    \]
    
\end{observation}

The last setup studied for this problem assumes that the information is stored in a set of sequences and the goal is to transmit a set $\cS$ of length-$\ell$ sequences of distributions over $\Sigma_q^w$. A \emph{single transmission} of a set of sequences $\mathcal{S}=\{\boldsymbol{s}_1,\dots,\boldsymbol{s}_k\}\subseteq(\Sigma_q^w)^{\ell}$ is a single transmission of a uniformly chosen sequence from $\cS$, i.e., a random sequence $\mathbf{X}^{\mathcal{S}}$ that is distributed like $\mathbf{X}^{\bfs_i}$ with probability $\frac{1}{k}$. An \emph{$n$-transmission} of a set of sequences $\mathcal{S}=\{\boldsymbol{s}_1,\dots,\boldsymbol{s}_k\}\subseteq(\Sigma_q^w)^{\ell}$ is a length-$n$ sequence of sequences $\mathbf{X}^{\cS}_n=(\mathbf{X}^{\cS,1},\dots,\mathbf{X}^{\cS,n})$ of independent single transmissions of the set $\cS$. 
A transmission $\mathbf{X}^{\mathcal{S}}$ of a set of sequences is a transmission of a uniformly chosen sequence with a label of the sent sequence, i.e., in every transmission one of the sequences is chosen uniformly and a transmission of the sequence is sent with a label so that the receiver can distinguish between the different sequences in the set. Similarly to the single strand case we have an $n$-transmission of a set of sequences $\mathcal{S}=(\boldsymbol{s}_1,\dots,\boldsymbol{s}_k)$ and refer to a sequence of sequences $\mathbf{X}^{\mathcal{S}}_n=(\mathbf{X}^{\mathcal{S},1},\dots,\mathbf{X}^{\mathcal{S},n})$  of length $n$ of independent single transmissions of the set $\mathcal{S}$.

The \emph{observed set} of the $n$-transmission $\boldsymbol{\mathbf{X}}^{\cS}_n$, denoted by $\Gamma(\boldsymbol{\mathbf{X}}^{\boldsymbol{s}}_n)$, is the set of all received single transmissions, i.e.,
$\Gamma(\boldsymbol{\mathbf{X}}^{\cS}_n)\triangleq\{\mathbf{X}^{\cS,i}\}_{i=1}^{n}\subseteq (\Sigma_q)^\ell.$
We denote by $\Gamma_j(\boldsymbol{\mathbf{X}}^{\boldsymbol{s}}_n)$ the subset of the observed set of transmissions with label $j$, i.e.,
$\Gamma_j(\boldsymbol{\mathbf{X}}^{\mathcal{S}}_n)\triangleq\{A\in\Gamma(\boldsymbol{\mathbf{X}}^{\mathcal{S}}_n) : A \mbox{ is labelled } j\}.$
An observed set \emph{recovers} the sequence $\boldsymbol{s}=(s_1,\dots,s_{\ell})\in(\Sigma_q^w)^{\ell}$ if for every $1\leq j\leq k$ the set $\Gamma_j(\boldsymbol{\mathbf{X}}^{\mathcal{S}}_n)$ recovers $\boldsymbol{s}_i$. Denote by $X_{\bfs_i}$ a random variable that indicates the number of single transmissions until it is possible to recover the sequence $\bfs_i$ by its observed set. From symmetry, we have that for any $\mathcal{S,S'}\subseteq(\Sigma_q^w)^{\ell}$ of size $k$ and $\bfs\in\cS,\bfs'\in\cS'$ it holds that $X_{\bfs}$ and $X_{\bfs'}$ have the same distribution and moreover every $\bfs_1,\bfs_2\in\cS$ have the same distribution as well. 
Therefore, given $\mathcal{S}=(\boldsymbol{s}_1,\dots,\boldsymbol{s}_k)$, we denote by $\text{RA}(\ell,w,q,k)$ the expected value to decode any sequence $\bfs_i\in\cS$, i.e., the random access expectation to decode $\bfs_i$, ${\text{RA}}(\ell,w,q,k)=\mathbf{E}[X_{{\boldsymbol{s}}_i}].$
Similarly to previous problems we can omit $q$.

\begin{problem}\label{prob:rand_acc}
Given $\ell,\omega,k\in\mathbf{N}$, find the value of $\text{RA}(\ell,w,k)$
\end{problem}

This value is given in the next theorem. A similar proof was given to another related problem in~\cite{CYB}.
\begin{theorem}
For every $\ell,w,k\in\mathbb{N}$ it holds that 
$$\text{RA}(\ell,w,k)=k{E}(\ell,w).$$
\end{theorem}
\begin{proof}
    Let $\mathcal{S}=(\boldsymbol{s}_1,\dots,\boldsymbol{s}_k)$ be an arbitrary set of sequences of size $k$.
    Let $X_{\bfs}$ be the random variable of $\bfs=\bfs_1$ for recovering $\bfs$ only with transmissions of $\bfs$, i.e.,
    \[
    {E}(\ell,w)=\mathbf{E}[X_{\bfs}].
    \]
    Let $X_{\bfs_1}$ be the random variable of $\bfs_1\in\cS$ for recovering $\bfs_1$ with transmissions from a set of size $k$ of sequences, i.e.,
    \[
    \text{RA}(\ell,w,k)=\mathbf{E}[X_{\bfs_1}].
    \]
    Using the tail sum formula for expected value we have that
    \[
    \mathbf{E}[{\mathbf{X}_{\boldsymbol{s}_1}}]=\sum_{i=0}^{\infty}{\mathbf{P}}[{\mathbf{X}_{\boldsymbol{s}_1}}>i].
    \]
    After $i$ transmissions it holds that $0\leq|\Gamma_1(\boldsymbol{\mathbf{X}}^{\mathcal{S}}_n)|\leq i$ summing over the different values of $\Gamma_1(\boldsymbol{\mathbf{X}}^{\mathcal{S}}_n)$ results for every $i$ that
    \[
    \mathbf{P}[{\mathbf{X}_{\boldsymbol{s}_1}}>i]=\sum_{j=0}^{i}\mathbf{P}[{\mathbf{X}_{\boldsymbol{s}_1}}>i\mid\;|\Gamma_1(\boldsymbol{\mathbf{X}}^{\mathcal{S}}_n)|=j]\mathbf{P}[\Gamma_1(\boldsymbol{\mathbf{X}}^{\mathcal{S}}_n)|=j].
    \]
    Since every transmission with label $1$ is distributed the same like $\bfs$ we have that
    \[
    \mathbf{P}[{\mathbf{X}_{\boldsymbol{s}_1}}>i\mid\;|\Gamma_1(\boldsymbol{\mathbf{X}}^{\mathcal{S}}_n)|=j]=\mathbf{P}[{\mathbf{X}_{\boldsymbol{s}}}>j].
    \]
    Since the sequence is chosen uniformly we have that
    \[
    \mathbf{P}[\Gamma_1(\boldsymbol{\mathbf{X}}^{\mathcal{S}}_n)|=j]=\binom{i}{j}(\frac{1}{k})^{j}(\frac{k-1}{k})^{i-j}.
    \]
    Therefore, it holds that
    \begin{align*}
    \mathbf{E}[{\mathbf{X}_{\boldsymbol{s}_1}}]=&\sum_{i=0}^{\infty}{\mathbf{P}}[{\mathbf{X}_{\boldsymbol{s}_1}}>i]
    =\sum_{i=0}^{\infty}\sum_{j=0}^{i}\mathbf{P}[{\mathbf{X}_{\boldsymbol{s}_1}}>i\mid|\Gamma_1(\boldsymbol{\mathbf{X}}^{\mathcal{S}}_n)|=j]\mathbf{P}[\Gamma_1(\boldsymbol{\mathbf{X}}^{\mathcal{S}}_n)|=j]\\
    =&\sum_{i=0}^{\infty}\sum_{j=0}^{i}\mathbf{P}[{\mathbf{X}_{\boldsymbol{s}}}>j]\binom{i}{j}(\frac{1}{k})^{j}(\frac{k-1}{k})^{i-j}.
    \end{align*}
    By changing the order of summation we get
    \begin{align*}
    \mathbf{E}[{\mathbf{X}_{\boldsymbol{s}_1}}]=&\sum_{j=0}^{\infty}\sum_{i=j}^{\infty}\mathbf{P}[{\mathbf{X}_{\boldsymbol{s}}}>j]\binom{i}{j}(\frac{1}{k})^{j}(\frac{k-1}{k})^{i-j}
    =\sum_{j=0}^{\infty}\mathbf{P}[{\mathbf{X}_{\boldsymbol{s}}}>j](\frac{1}{k})^{j}\sum_{i=j}^{\infty}\binom{i}{j}(\frac{k-1}{k})^{i-j}.
    \end{align*}
    It is known that $    \sum_{i=0}^{\infty}x^i=\frac{1}{1-x}$
    and by differentiating $j$ times we get
    \[
    \sum_{i=0}^{\infty}i(i-1)\cdots(i-j+1)x^{i-j}=j!\frac{1}{(1-x)^{j+1}},
    \]
    and hence,
    $$\sum_{i=0}^{\infty}\binom{i}{j}x^{i-j}=\frac{1}{(1-x)^{j+1}}.$$
    One can observe that $$\sum_{i=0}^{\infty}\binom{i}{j}x^{i-j}=\sum_{i=j}^{\infty}\binom{i}{j}x^{i-j},$$ so by setting $x=\frac{k-1}{k}$ we get 
    \begin{align*}
    \mathbf{E}[{\mathbf{X}_{\boldsymbol{s}_1}}]=&\sum_{j=0}^{\infty}\mathbf{P}[{\mathbf{X}_{\boldsymbol{s}}}>j](\frac{1}{k})^{j}\sum_{i=j}^{\infty}\binom{i}{j}(\frac{k-1}{k})^{i-j}
    =\sum_{j=0}^{\infty}\mathbf{P}[{\mathbf{X}_{\boldsymbol{s}}}>j](\frac{1}{k})^{j}\frac{1}{(1-\frac{k-1}{k})^{j+1}}
    =k\sum_{j=0}^{\infty}\mathbf{P}[{\mathbf{X}_{\boldsymbol{s}}}>j] =k\mathbf{E}[{\mathbf{X}_{\boldsymbol{s}}}].
    \end{align*}
\end{proof}

In the following section we study composite alphabets of fixed size. In the previous problem, we studied the use of $\omega$-composite symbols, and in this section we examine all the possible composite symbols. The goal is to find optimal sets of composite symbols to create an optimal alphabet.
\section{Optimal Composite Alphabets}\label{sec:defs}

Every $\bfrh\in \Sigmaexp$ is referred to as a \emph{composite symbol}. A \emph{single transmission} of a composite symbol $\bfrh\in\Sigmaexp$ is a random variable  $X^{\bfrh}$ over the distribution $\bfrh$, and the single transmission refers to transmitting a single symbol over the composite DNA channel according to the distribution $\bfrh$. An \emph{$n$-transmission} is a vector of $n$ independent and identical random variables 
$\mathbf{X}^{\bfrh}_n=(X^{\bfrh}_1,\dots,X^{\bfrh}_n),$
each is over the distribution of the composite symbol $\bfrh\in\Sigmaexp$. It is also referred to as a sequence of $n$ different and independent single transmissions of $\bfrh$, and $n$ denotes the \emph{size} of the transmission. 

Given a symbol $i\in\Sigma_q$, denote by ${\#_i}(\mathbf{X}_n^{\bfrh})$ the number of occurrences of $i$ in the $n$-transmission vector $\mathbf{X}^{\bfrh}_n$. Denote by $\Theta(\mathbf{X}^{\bfrh}_n)$ the \emph{observed distribution} of the $n$-transmission $\mathbf{X}^{\bfrh}_n$, i.e.,
$\Theta(\mathbf{X}^{\bfrh}_n)= \left(\frac{{\#_1}(\mathbf{X}_n^{\bfrh})}{n}
,\dots,
\frac{{\#_q}(\mathbf{X}_n^{\bfrh})}{n}\right).$
An observed distribution refers to the received information in the receiver side, which is the information that is used to decode the composite symbol that was used for the $n$-transmission.

\begin{example}
Using DNA as our alphabet of size 4, we let $\Sigma_4=\{\sfA(1),\sfC(2),\sfG(3),\sfT(4)\}.$
The composite symbol
$\bfrh=(\frac{1}{2}, \frac{1}{2}, 0, 0)$ 
belongs to the DNA composite alphabet. Throughout this section we use colored figures to illustrate the examples. An illustration for the composite symbol $\bfrh$ is shown in Fig.
\ref{fig:rho_example}.

\begin{figure}[H]
    \centering
    \includegraphics[width=0.25\linewidth]{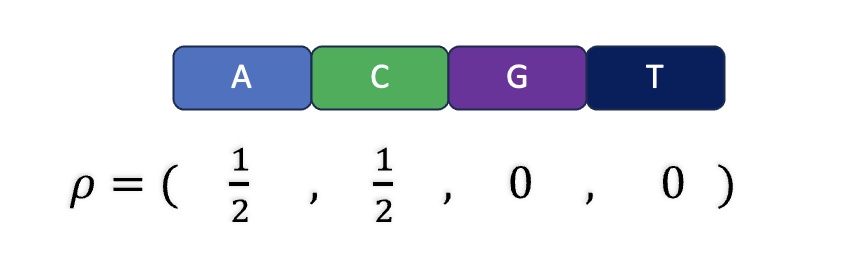}
    \caption{The composite symbol $\rho$}
    \label{fig:rho_example}
\end{figure}

A single transmission $\mathbf{X}^{\bfrh}$ is a random variable that with probability $\frac{1}{2}, \frac{1}{2}, 0, 0$ is equal to $\sfA,\sfC,\sfG,\sfT$, respectively.
A possible 5-transmission $\mathbf{X}^{\bfrh}_5$ is shown in Fig. \ref{fig:trans_5_example}.

\begin{figure}[H]
    \centering
    \includegraphics[width=0.25\linewidth]{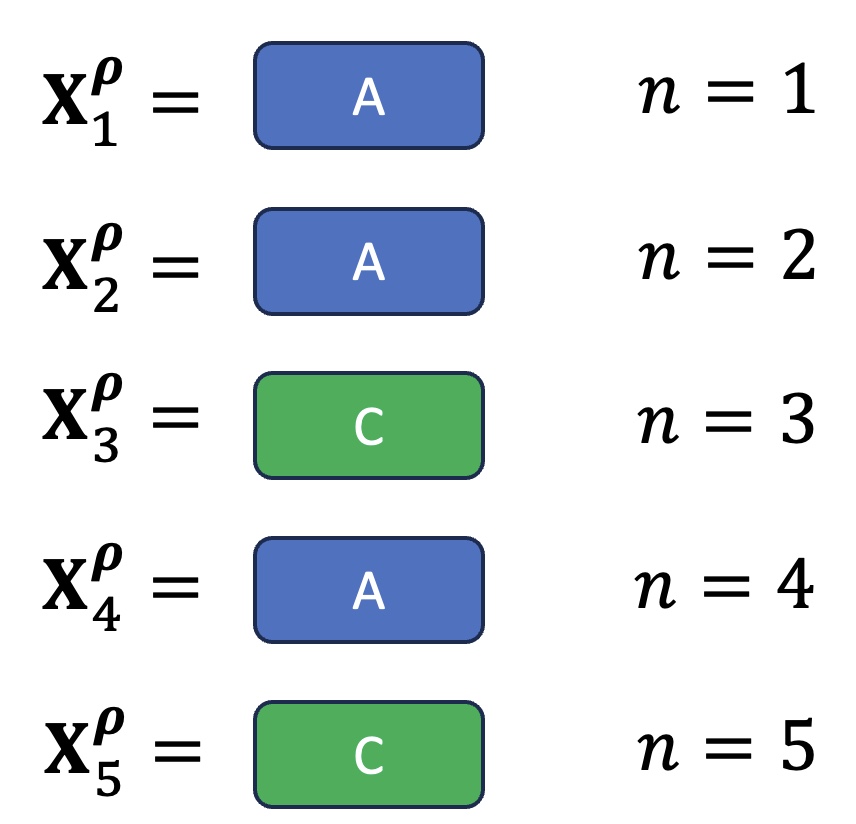}
    \caption{A 5-transmission for $\rho$}
    \label{fig:trans_5_example}
\end{figure}

There are $3$ $\sfA$'S in the transmission and $2$ $\sfC$'S. Therefore, the observed distribution of the $5$-transmission is
\[
\Theta(\mathbf{X}^{\bfrh}_5)=\left(\frac{\textcolor{RoyalBlue}{3}}{5},\frac{\textcolor{Green}{2}}{5},0,0\right).
\]

Denote by $\Omega_n^q$ the set of all possible observed distributions of $n$-transmissions over the composite DNA channel, i.e., 
$$\Omega_n^q=\big\{(\frac{i_1}{n},\dots,\frac{i_\AlphabetSize}{n}) : i_1,\dots,i_\AlphabetSize\in\mathbb{N}\mbox{ and } \sum_{j=1}^\AlphabetSize{i_j}=n\big\}.$$
\end{example}

Denote by $\Omega^q$ the union of all possible observed distributions of all $n$-transmissions over the composite DNA channel for every $n$, i.e., $
\Omega^q=\bigcup_{n\in\N}{\Omega_n^q}$.
An \emph{$(m,q)$-composite code} $\mathcal{C}$ is a set of $m$ composite symbols in $\Sigmaexp$, i.e., $\mathcal{C}=\{\bfc_1,\dots,\bfc_m\}\subseteq\Sigmaexp$. Denote by $\mathcal{C}_q^m$ all possible $(m,q)$-composite codes. Our goal is to study how to choose the distributions from $\Sigmaexp$ such that given the observed distribution, the decoding success probability will be maximized. 
Namely, given:
\begin{itemize}
    \item an $(m,q)$-composite code $\mathcal{C}$.
    \item the size of the transmission, $n$.
    \item a decoder function $\mathcal{D}:\Omega_n^q\times \mathbb{N}\rightarrow \mathcal{C}$, from all the observed distributions of $n$-transmissions and the transmission size to a composite symbol from the composite code $\mathcal{C}$,
\end{itemize}
we define the \emph{decoding success probability} of the composite symbol $\boldsymbol{c}\in\mathcal{C}$ to be
\[
\psucc(\cC,n,\mathcal{D},\bfc)\triangleq\mathbf{P}[\mathcal{D}(\Theta(\mathbf{X}^{\bfc}_n))=\bfc],
\]
i.e., the probability that an observed distribution of an $n$-transmission of  $\boldsymbol{c}$ is decoded to the transmitted composite symbol. There are two figure of merits we seek to optimize:
\begin{itemize}
    \item The minimum decoding success probability, $$\fmin(\mathcal{C},n,\mathcal{D})\triangleq\min_{\boldsymbol{c}\in\mathcal{C}}\psucc(\mathcal{C},n,\mathcal{D},\boldsymbol{c}).$$
    \item The average decoding success probability, $$\favg(\mathcal{C},n,\mathcal{D})\triangleq\frac{1}{m}\sum_{\boldsymbol{c}\in\mathcal{C}}\psucc(\mathcal{C},n,\mathcal{D},\boldsymbol{c}).$$
\end{itemize}
Our goal is to study the following optimization functions.
\begin{problem}
Given $m,q,$ and $n$, find an optimal $(m,q)$-composite code $\mathcal{C}$ and a decoder\footnote{The set of a code and a decoder might not be unique} that optimize ${f\in\{\fmin, \favg\}}$ and this optimization value. That is, find the values of
\begin{enumerate}
\item $\optmin^n(m,q)\triangleq\arg\max_{\mathcal{C}\in\mathcal{C}_q^m,\mathcal{D}}\fmin(\mathcal{C},n,\mathcal{D})$,
\item $\optavg^n(m,q)\triangleq\arg\max_{\mathcal{C}\in\mathcal{C}_q^m,\mathcal{D}}\favg(\mathcal{C},n,\mathcal{D})$,
\item $\optminprob^n(m,q)=
\max_{\mathcal{C}\in\mathcal{C}_q^m,\mathcal{D}}\fmin(\mathcal{C},n,\mathcal{D})$,
\item $\optavgprob^n(m,q)=
\max_{\mathcal{C}\in\mathcal{C}_q^m,\mathcal{D}}\favg(\mathcal{C},n,\mathcal{D})$.
\end{enumerate}
\end{problem}

In the following sections we analyze different decoders for our model. We specify their optimality or rather the lack of it. Next we find optimal composite alphabets and their properties using the MLD. 

\section{Analysis of the Maximum Likelihood Decoder}\label{sec:MLD}

Let us present the \emph{Maximum Likelihood Decoder} (\emph{MLD}), denoted by $\Dmld$. The MLD is defined over any $(m,q)$-composite code $\mathcal{C}$ as follows. For every $\bfth\in\Omega_n^q$ and $n>0$
\[
    \Dmld(\bfth, n)= \mbox{argmax}_{\boldsymbol{c}\in\mathcal{C}}{\mathbf{P}[\Theta(\mathbf{X}^{\boldsymbol{c}}_n)=\bfth]},
\]
and if there is more than one composite symbol that maximizes this probability, the decoder chooses the first one in lexicographic order of their distributions.

The following lemmas show that the MLD does not change with respect to the number of samples $n$. This property will help us to omit the number of transmissions from the MLD.

\begin{lemma}\label{lm:MLD_n_free_equality}
For any $(m,q)$-composite code $\mathcal{C}$, if $\bftau\in\Omega_{n_1}^q\cap\Omega_{n_2}^q$ then for every $\bfc_1,\bfc_2\in\cC$ it holds that,
$${\mathbf{P}[\Theta(\mathbf{X}^{\bfc_1}_{n_1})=\bftau]} \geq \mathbf{P}[\Theta(\mathbf{X}^{\bfc_2}_{n_1})=\bftau]$$ if and only if $$\mathbf{P}[\Theta(\mathbf{X}^{\bfc_1}_{n_2})=\bftau] \geq \mathbf{P}[\Theta(\mathbf{X}^{\bfc_2}_{n_2})=\bftau].$$

\end{lemma}
\begin{proof}
    Let $\mathcal{C}$ be an $(m,q)$-composite code and $\bftau\in \Omega_{n_1}^q\cap\Omega_{n_2}^q$.
    Denote $\mathcal{C}=\{\bfc_1,\dots,\bfc_m\}$, $\bfc_i=(c_{i,1},\dots,c_{i,q})$ and $\bftau=(\tau_1,\dots,\tau_q)$. Assume without loss of generality that $\Dmld(\bftau, n_1)=\bfc_1$, and we will prove that $\Dmld(\bftau, n_2)=\bfc_1$. 
    
    Let $\bfc_1,\bfc_2\in\cC$, denote $\bfc_i=(c_{i,1},\dots,c_{i,q})$. By definition for $i,j\in\{1,2\}$ it holds that
\[
    \mathbf{P}[\Theta(\mathbf{X}^{\bfc_i}_{n_j})=\bftau]=\binom{n_j}{\tau_1n_j,\dots,\tau_qn_j}\prod_{k=1}^q(c_{i,k})^{\tau_in_j}.
\]
If it holds that
\[
\mathbf{P}[\Theta(\mathbf{X}^{\bfc_1}_{n_1})=\bftau]\geq\mathbf{P}[\Theta(\mathbf{X}^{\bfc_2}_{n_1})=\bftau].
\]
then it holds that 
$$\binom{n_1}{\tau_1n_1,\dots,\tau_qn_1}\prod_{i=1}^q(c_{1,i})^{\tau_in_1}\geq \binom{n_1}{\tau_1n_1,\dots,\tau_qn_1}\prod_{i=1}^q(c_{2,i})^{\tau_in_1},$$
which implies that 
$$\prod_{i=1}^q(c_{1,i})^{\tau_in_1}\geq \prod_{i=1}^q(c_{2,i})^{\tau_in_1},$$
that is, 
$$\prod_{i=1}^q(c_{1,i})^{\tau_i}\geq \prod_{i=1}^q(c_{h,i})^{\tau_i}.$$
Following the same steps reversely with $n_2$ we get that
$$\binom{n_2}{\tau_1n_2,\dots,\tau_qn_2}\prod_{i=1}^q(c_{1,i})^{\tau_in_2}\geq
\binom{n_2}{\tau_1n_2,\dots,\tau_qn_2}\prod_{i=1}^q(c_{2,i})^{\tau_in_2}.$$
Therefore, 
\[
\mathbf{P}[\Theta(\mathbf{X}^{\bfc_1}_{n_2})=\bftau]\geq\mathbf{P}[\Theta(\mathbf{X}^{\bfc_2}_{n_2})=\bftau].
\]
\end{proof}

\begin{corollary}\label{lm:MLD_n_free}
For any $(m,q)$-composite code $\mathcal{C}$, if $\bftau\in\Omega_{n_1}^q\cap\Omega_{n_2}^q$ then it holds that
$$\mbox{argmax}_{\boldsymbol{c}\in\mathcal{C}}{\mathbf{P}[\Theta(\mathbf{X}^{\boldsymbol{c}}_{n_1})=\bfth]}=\mbox{argmax}_{\boldsymbol{c}\in\mathcal{C}}{\mathbf{P}[\Theta(\mathbf{X}^{\boldsymbol{c}}_{n_2})=\bfth]}.$$
\end{corollary}

dAs a result of \autoref{lm:MLD_n_free}, we deduce that the MLD does not depend on $n$\footnote{One can note that the claim is not fully accurate since not all the distributions are in the domain of MLD for every $n$. However, for every distribution that exists in the domain of $n_1,n_2$ the MLD returns the same symbol.}. Therefore, from now on we will use $\Dmld(\bfth)$ to describe the MLD for any $n$ and for any $\bfth\in\Omega^q$.

For any $(m,q)$-composite code $\mathcal{C}$, denote by $\mbox{DR}_{\mathcal{D}}^n(\boldsymbol{c})$ the \emph{decoding region} (DR) of $\bfc\in\cC$ with $n$-transmissions using $\mathcal{D}$, which is the set of all distributions that are decoded to $\boldsymbol{c}$, i.e., $$\text{DR}_{\mathcal{D}}^n(\boldsymbol{c})\triangleq\left\{\bfth \in \Omega_n^q:\mathcal{D}(\bfth, n)=\boldsymbol{c}\right\}.$$
Since $\mathcal{D}$ is a function from $\Omega_n^q$ it holds that $
\bigcup_{\boldsymbol{c}\in\mathcal{C}}\text{DR}_{\mathcal{D}}^n(\boldsymbol{c})=\Omega_n^q$, and for every $\bfc_1,\bfc_2\in\mathcal{C}$ such that $\bfc_1\neq\bfc_2$ it holds that
$\text{DR}_{\mathcal{D}}^n(\bfc_1)\cap\text{DR}_{\mathcal{D}}^n(\bfc_2)=\phi$. Since $\Dmld$ does not depend on $n$ we denote for every $\boldsymbol{c}\in\mathcal{C}$, $
\text{DR}^n_{\Dmld}(\boldsymbol{c})\triangleq\drmld(\boldsymbol{c})$.

We start with a first property on the value of $\favg(\mathcal{C},n,\mathcal{D})$.
\begin{lemma}\label{lm:split_avg_to_sum}
    For any $(m,q)$-composite code $\mathcal{C}$, decoder $\mathcal{D}$, and $n\in\mathbb{N}$ it holds that $\favg(\mathcal{C},n,\mathcal{D})=\frac{1}{m}\sum_{\bfth\in\Omega_n^q}\mathbf{P}[(\Theta(\mathbf{X}^{\mathcal{D}(\bfth,n)}_n)=\bfth].$
\end{lemma}
\begin{proof}
 Let $\mathcal{C}$ be an $(m,q)$-composite code and $n\in\mathbb{N}$. 
 By definition, $\favg(\mathcal{C},n,\mathcal{D})$ equals to 
\begin{align*}
=&\frac{1}{m}\sum_{\boldsymbol{c}\in\mathcal{C}}\psucc(\mathcal{C},n,\mathcal{D},\boldsymbol{c})=\frac{1}{m}\sum_{\boldsymbol{c}\in\mathcal{C}}\mathbf{P}[\mathcal{D}(\Theta(\mathbf{X}^{\boldsymbol{c}}_n),n)=\boldsymbol{c}] \\
=&\frac{1}{m}\sum_{\boldsymbol{c}\in\mathcal{C}}\sum_{\bfth\in \text{DR}^n_{\mathcal{D}}(\boldsymbol{c})}\mathbf{P}[\Theta(\mathbf{X}^{\boldsymbol{c}}_n)=\bfth]. 
\end{align*}
Since the DRs are distinct and their union is $\Omega_n^q$, we get that $
\favg(\mathcal{C},n,\mathcal{D})=\frac{1}{m}\sum_{\bfth\in\Omega_n^q}\mathbf{P}[(\Theta(\mathbf{X}^{\mathcal{D}(\bfth,n)}_n)=\bfth]$.
\end{proof}
The next theorem proves the MLD's optimality for $\favg$.
\begin{theorem}\label{th:MLD_optimal_avg}
For any $(m,q)$-composite code $\mathcal{C}$, $\Dmld$ maximizes $\favg$, i.e., for every decoder $\mathcal{D}$ and $n\in\mathbb{N}$
\[
\favg(\mathcal{C},n,\mathcal{D})\leq
\favg(\mathcal{C},n,\Dmld).
\]
\end{theorem}
\begin{proof}
According to Lemma \ref{lm:split_avg_to_sum} it holds that
$\favg(\mathcal{C},n,\mathcal{D})=\frac{1}{m}\sum_{\bfth\in\Omega_n^q}\mathbf{P}[(\Theta(\mathbf{X}^{\mathcal{D}(\bfth,n)}_n)=\bfth]$. 
By the definition of the MLD it holds that for any $\bfth\in\Omega_n^q$ for every $\boldsymbol{c}\in\mathcal{C}$ it holds that
 $
 \mathbf{P}[\Theta(\mathbf{X}^{\boldsymbol{c}}_n)=\bfth]\leq\mathbf{P}[\Theta(\mathbf{X}^{\Dmld(\bfth)})=\bfth].$ Since ${\mathcal{D}(\bfth,n)}\in\mathcal{C}$ it holds that $\frac{1}{m}\sum_{\bfth\in\Omega_n^q}\mathbf{P}[(\Theta(\mathbf{X}^{\mathcal{D}(\bfth,n)}_n)=\bfth]\leq$
$\frac{1}{m}\sum_{\bfth\in\Omega_n^q}\mathbf{P}[(\Theta(\mathbf{X}^{\Dmld(\bfth)}_n)
    =\bfth]=\favg(\mathcal{C},n,\Dmld).$
 \end{proof}
The next theorem disproves the MLD's optimality for $\fmin$.
\begin{theorem}\label{th:MLD_not_optimal_min}
The MLD is not optimal for $\fmin$, i.e., there exist an $(m,q)$-composite code $\mathcal{C}$, a decoder $\mathcal{D}$ and $n\in\mathbb{N}$ such that $\fmin(\mathcal{C},n,\mathcal{D}_{\mbox{MLD}})<\fmin(\mathcal{C},n,\mathcal{D})$.
\end{theorem}
\begin{proof}
We prove using an example with a $(3,2)$-composite code $\mathcal{C}=\{\bfc_1,\bfc_2,\bfc_3\}$ and $n=10$,
\[
\bfc_1=(\frac{4}{10},\frac{6}{10}),\bfc_2=(\frac{1}{2},\frac{1}{2}),\bfc_3=(\frac{6}{10},\frac{4}{10}).
\]
One can calculate that,
\begin{gather*}
  \drmld(\bfc_1)=\{(x,1-x)\in\alldist : x < \frac{\log\frac{10}{12}}{\log\frac{4}{6}}\ \approx 0.449\},\\
  \drmld(\bfc_3)=\{(1-y,y)\in\alldist : y < \frac{\log\frac{10}{12}}{\log\frac{4}{6}}\ \approx 0.449\},\\
  \drmld(\bfc_2)=\alldist\setminus(\drmld(\bfc_1)\cup\drmld(\bfc_3)).
\end{gather*}
Therefore,
\begin{gather*}
\mathbf{P}[\Dmld(\Theta(\mathbf{X}^{\bfc_1}_{10}))=\bfc_1]=\sum_{i=0}^4{\binom{10}{i}}(\frac{4}{10})^i(\frac{6}{10})^{10-i}\approx0.633,\\
  \mathbf{P}[\Dmld(\Theta(\mathbf{X}^{\bfc_3}_{10}))=\bfc_3]=\sum_{i=6}^{10}{\binom{10}{i}}(\frac{6}{10})^i(\frac{4}{10})^{10-i}\approx0.633,\\
  \mathbf{P}[\Dmld(\Theta(\mathbf{X}^{\bfc_2}_{10}))=\bfc_2]=\binom{10}{5}(\frac{1}{2})^{10}\approx0.246.
\end{gather*}
Thus, we get that
\[
\fmin(\mathcal{C},n,\Dmld)=\mathbf{P}[\Dmld(\Theta(\mathbf{X}^{\bfc_2}_{10}))=\bfc_2]\approx0.246.
\]
Define $\mathcal{D}'$ by
\[
   \mathcal{D}'(\boldsymbol{c}, 10)=
    \begin{cases}
    \mathcal{D}_{\mbox{MLD}}(\boldsymbol{c}),& \text{if } \boldsymbol{c}\neq(0,1)\\
    \bfc_2             & \boldsymbol{c}=(0,1)
    \end{cases}.
\]
Now we have,
\begin{gather*}
\mathbf{P}[\mathcal{D}'(\Theta(\mathbf{X}^{\bfc_1}_{10}), 10)=\bfc_1]=\sum_{i=1}^4{\binom{10}{i}}(\frac{4}{10})^i(\frac{6}{10})^{10-i}\approx0.627,\\
  \mathbf{P}[\mathcal{D}'(\Theta(\mathbf{X}^{\bfc_3}_{10}), 10)=\bfc_3]=\sum_{i=6}^{10}{\binom{10}{i}}(\frac{6}{10})^i(\frac{4}{10})^{10-i}\approx0.633,\\
  \mathbf{P}[\mathcal{D}'(\Theta(\mathbf{X}^{\bfc_2}_{10}), 10)=\bfc_2]=\binom{10}{5}(\frac{1}{2})^{10}+(\frac{1}{2})^{10}\approx0.247.
\end{gather*}
Thus, we get that
\begin{gather*}
\fmin(\mathcal{C},n,\mathcal{D}')= \mathbf{P}[\mathcal{D}'(\Theta(\mathbf{X}^{\bfc_2}_{10}), 10)=\bfc_2]>\\
\mathbf{P}[\Dmld(\Theta(\mathbf{X}^{\bfc_2}_{10}))=\bfc_2]=\fmin(\mathcal{C},n,\Dmld).
\end{gather*}
\end{proof}

Using \autoref{th:MLD_optimal_avg} for every $(m,q)$-composite code $\mathcal{C}$, $\Dmld$ is optimal to optimize $\favg$. Hence, for any optimal $(m,q)$-composite code $\mathcal{C}$ and decoder $\mathcal{D}$ in $\optavg^n(q,m)$, it holds that $(\mathcal{C},\Dmld)\in\optavg^n(q,m)$.
As a result, from now on in this paper we will denote $
\optavg^n(q,m)\triangleq
\max_{\mathcal{C}\in\mathcal{C}_q^m}\favg(\mathcal{C},n,\Dmld)$.
Although the MLD is not optimal to optimize $\fmin$, as shown in \autoref{th:MLD_not_optimal_min}, we will denote similarly $\optmin^n(q,m)\triangleq
\max_{\mathcal{C}\in\mathcal{C}_q^m}\fmin(\mathcal{C},n,\Dmld)$. Thus, our goal is to study composite codes which optimize these success probabilities when using the MLD.

\section{Basic Optimal Alphabets}\label{sec:basic}
By the properties of the probability function, for every $(m,q)$-composite code $\mathcal{C}$, $n\in\mathbb{N}$ and a decoder $\mathcal{D}$ it holds that $\fmin(\mathcal{C},n,\mathcal{D}),\favg(\mathcal{C},n,\mathcal{D})\leq 1$.
We denote the $\emph{support}$ of a composite symbol $\bfrh=(\rho_1,\dots,\rho_q)$ to be all the indices with positive probability, i.e.,
$\mbox{supp}(\bfrh)\triangleq\left\{i:\rho_i>0\right\}.$ A composite symbol $\boldsymbol{c}$ with $\mbox{supp}(\boldsymbol{c})$ of size $1$ is referred as a \emph{base composite symbol}. Denote by $e_{q,i}$ the base composite symbol over $\Sigma_q$, with $1$ in the $i$-th index, $\bfe_{q,i}\triangleq(\dots,0,1,0,\dots)$. The set of all base composite symbols over $\Sigma_q$ is referred as the \emph{base composite alphabet} and this set is denoted by $$E_q\triangleq\{\bfe_{q,1},\dots,\bfe_{q,i},\dots,\bfe_{q,q}\}.$$

Given a composite symbol $\boldsymbol{c}\in\Sigmaexp$ and $n\in\mathbb{N}$, denote by $\imn(\boldsymbol{c})$ all the observed distributions from $\bfrh$ with positive probability, i.e., 
\[
\imn(\boldsymbol{c})=\left\{\bfth\in\Omega_n^q : {\mathbf{P}[(\Theta(\mathbf{X}^{\boldsymbol{c}}_{n}))=\bfth]}>0  \right\}.
\]
It holds that $ \mathbf{P}[(\Theta(\mathbf{X}^{\boldsymbol{c}}_{n}))\in\imn(\boldsymbol{c})]=1$.
Hence, for any $(m,q)$-composite code $\mathcal{C}$ that contains $\boldsymbol{c}$ and a decoder $\mathcal{D}$ it holds that 
\[
\psucc(\mathcal{C},n,\mathcal{D},\boldsymbol{c})=\mathbf{P}[(\Theta(\mathbf{X}^{\boldsymbol{c}}_{n}))\in\imn(\boldsymbol{c})\cap\text{DR}_{\mathcal{D}}^n(\boldsymbol{c})].
\]
The following lemma proves that the support of an observed distribution can only be a subset of the original symbol's support.
\begin{lemma}\label{lm:supp_subset_orig}
Given a composite symbol $\boldsymbol{c}\in\Sigmaexp$ and $n\in\mathbb{N}$, for every $\bfth\in\imn(\boldsymbol{c})$ it holds that 
\[
\mbox{supp}(\bfth)\subseteq\mbox{supp}(\boldsymbol{c}).
\]
\end{lemma}
\begin{proof}
    Let $\boldsymbol{c}\in\Sigmaexp$ be a composite symbol, $n\in\mathbb{N}$ and $\bfth\in\imn(\boldsymbol{c})$.
    If $i\in\mbox{supp}(\bfth)$ then there exists an $n-$transmission $\boldsymbol{a}=(a_1,\dots,a_n)$ such that
    \begin{itemize}
        \item $\Theta(\boldsymbol{a})=\bfth$,
        \item There exists $j$ such that $a_j=i$, since $i\in\mbox{supp}(\bfth)$,
        \item $\mathbf{P}[(\mathbf{X}^{\boldsymbol{c}}_{n})= \boldsymbol{a}]>0$, since $\bfth\in\imn(\boldsymbol{c})$.
    \end{itemize}
    Therefore $c_i>0$ which implies that $i\in\mbox{supp}(\boldsymbol{c})$.
\end{proof}

Now we are interested in codes that one can decode all the symbols with success decoding probability of 1. The following theorem provides the required and sufficient properties for a code to acquire this property.

\begin{theorem}\label{th:perfect_decode_distinct_support}
For every $(m,q)$-composite code $\mathcal{C}$ the following conditions are equivalent,
\begin{enumerate}
    \item For every $\bfc_1,\bfc_2\in\mathcal{C}$ such that $\bfc_1\neq\bfc_2$ it holds that $\mbox{supp}(\bfc_1)\cap\mbox{supp}(\bfc_2)=\phi$.
    \item For every $n\in\mathbb{N}$ and $\boldsymbol{c}\in\mathcal{C}$ it holds that $\psucc(\mathcal{C},n,\Dmld,\boldsymbol{c})=1$.
    \end{enumerate}
\end{theorem}

\begin{proof}
    $1 \Rightarrow 2$.
    Let $\mathcal{C}$ be an $(m,q)$-composite code that fulfills $(1)$. 
    Let $\bfc_1,\bfc_2\in\mathcal{C}$ be such that $\bfc_1\neq\bfc_2$. We know that $\mbox{supp} (\bfc_1)\cap\mbox{supp}(\bfc_2)=\phi$, therefore  $\imn(\bfc_1)\cap\imn(\bfc_2)=\phi$ using Lemma \autoref{lm:supp_subset_orig}.
    For any $\boldsymbol{c}\in\mathcal{C}$ it holds that for every $\bfth\in\imn(\boldsymbol{c})$
    \[
    \mathbf{P}[\Theta(\mathbf{X}^{\boldsymbol{c}}_{n})= \bfth]>0,
    \]
    and for any $\boldsymbol{c'}\in\mathcal{C}$ such that $\boldsymbol{c'}\neq\boldsymbol{c}$ we know that $\imn(\boldsymbol{c})\cap\imn(\boldsymbol{c'})=\phi$, therefore 
    \[
    \mathbf{P}[\theta(\mathbf{X}^{\boldsymbol{c'}}_{n})= \bfth]=0.
    \]
    Hence by the definition of the MLD we get for every $\boldsymbol{c}\in\mathcal{C}$
    \[
    \imn(\boldsymbol{c})\subseteq\drmld(\boldsymbol{c}).
    \]
    Thus, we get
    \[
    \psucc(\mathcal{C},n,\Dmld,\boldsymbol{c})=\mathbf{P}[\Theta(\mathbf{X}^{\boldsymbol{c}}_{n})\in\imn(\boldsymbol{c})\cap\drmld(\boldsymbol{c})]
    \]
    \[
    =\mathbf{P}[(\Theta(\mathbf{X}^{\boldsymbol{c}}_{n}))\in\imn(\boldsymbol{c}))]=1
    \]
    $2 \Rightarrow 1$
    Let $\mathcal{C}$ be an $(m,q)$-composite code that does not fulfill $(1)$ and let $\bfc_1,\bfc_2\in\mathcal{C}$ $\bfc_1\neq\bfc_2$ be such that $\mbox{supp}(\bfc_1)\cap\mbox{supp}(\bfc_2)\neq\phi$,
    w.l.o.g assume $1\in\mbox{supp}(\bfc_1)\cap\mbox{supp}(\bfc_2)$.
    Thus, we get that for every $n\in\mathbb{N}$
    \[
    \bfe_{q,1}\in\imn(\bfc_1)\cap\imn(\bfc_2).
    \]
    Let $n\in\mathbb{N}$, we know that
    \[
    \drmld(\bfc_1)\cap\drmld(\bfc_2)=\phi
    \]
    w.l.o.g $e_{q,1}\notin\drmld(\bfc_1)$ results with
    \[
    \imn(\bfc_1)\cap\drmld(\bfc_1)\subsetneqq \imn(\bfc_1).
    \]
    So,
    \[
    \psucc(\mathcal{C},n,\Dmld,\bfc_1)=\mathbf{P}[(\Theta(\mathbf{X}^{\bfc_1}_{n}))\in\imn(\bfc_1)\cap\drmld(\bfc_1)]
    \]
    \[
    \leq\mathbf{P}[(\Theta(\mathbf{X}^{\bfc_1}_{n}))\in\imn(\bfc_1))]-\mathbf{P}[(\Theta(\mathbf{X}^{\bfc_1}_{n}))=e_{q,1}]<1.
    \]
    Since
    \[
    \mathbf{P}[(\Theta(\mathbf{X}^{\bfc_1}_{n}))=e_{q,1}]>0.
    \]
    Thus, we get that $\mathcal{C}$ does not fulfill $(2)$
    
\end{proof}

An $(m,q)$-composite code is called a \emph{distinct support code} if for every $\bfc_1,\bfc_2\in\mathcal{C}$ such that $\bfc_1\neq\bfc_2$ it holds that $\mbox{supp}(\bfc_1)\cap\mbox{supp}(\bfc_2)=\phi$. Denote by $\mathcal{DS}^m_q$ the set of all $(m,q)$ distinct support composite codes. One can verify that $\mathcal{DS}^1_q=\Sigmaexp$ and  $\mathcal{DS}^q_q=\{E_q\}.$

The following results are deduced from \autoref{th:perfect_decode_distinct_support}. For every $m,q,n\in\mathbb{N}$ such that $m\leq q$ it holds that

\begin{itemize}
  \item $\optmin(m, q)=\mathcal{DS}^m_q$,
  \item $\optminprob(m, q)=1$,
  \item $\optavg(m, q)=\mathcal{DS}^m_q$,
  \item $\optavgprob(m, q)=1$.
\end{itemize}

The following lemmas prove the surmise that optimal codes should always contain the base composite alphabets. The full proof is shown in Appendix~\ref{app:A}.
\begin{restatable}{lemma}{lmaperfectbaseLm}\label{lm:perfect_decode_base_composite}
    For any $(m,q)$-composite code $\mathcal{C}$ if $\bfe_{q,i}\in\mathcal{C}$ it holds that for any $n\in\mathbb{N}$, $\psucc(\mathcal{C},n,\Dmld,\bfe_{q,i})=1.$
\end{restatable}

\begin{restatable}{lemma}{samewithallbaseLemma}\label{lm:same_with_all_base}
    For any $(m,q)$-composite code $\mathcal{C}$ with $m\geq q$, there exists an $(m,q)$-composite code $\mathcal{C}'$ such that
    \begin{itemize}
        \item $E_q\subseteq\mathcal{C}'$,
        \item $\favg(\mathcal{C'},n,\Dmld)\geq\favg(\mathcal{C},n,\Dmld)$,
        \item $\fmin(\mathcal{C'},n,\Dmld)\geq\fmin(\mathcal{C},n,\Dmld)$.
    \end{itemize}
\end{restatable}

The following claim justifies  narrowing the search for optimal codes only to codes that contain the base composite alphabet.
Denote by $\codeconbase$ the set of all $(m,q)$-composite codes that contain $E_q$, i.e.,
\[
\codeconbase=\big\{\cC\in\mathcal{C}_q^m : E_q\subseteq\cC\big\}.
\]
\begin{claim}\label{cl:search_only_with_base}
For any $q,m,n\in\mathbb{N}$ such that $m\geq q$ it holds that $\optavg^n(q,m)=\phi$ if and only if  $\arg\max_{\mathcal{C}\in\codeconbase}\favg(\mathcal{C},n,\Dmld)=\phi$. Furthermore, $\optmin^n(q,m)=\phi$ if and only if  $\arg\max_{\mathcal{C}\in\codeconbase}\fmin(\mathcal{C},n,\Dmld)=\phi$.
\end{claim}
\begin{proof}
    We will prove the lemma for $\optavg$, however the proof for $\optmin$ is identical.
    
    $\Rightarrow$ Assume $\arg\max_{\mathcal{C}\in\codeconbase}\favg(\mathcal{C},n,\Dmld)\neq\phi$, and let $\mathcal{C}\in\arg\max_{\mathcal{C}\in\codeconbase}\favg(\mathcal{C},n,\Dmld)$. If $\mathcal{C}\notin\optavg^n(q,m)$ then there exists an $(m,q)$-composite code $\mathcal{C}'$ such that
    \[\favg(\mathcal{C}',n,\Dmld)>\favg(\mathcal{C},n,\Dmld).
    \]
    From Lemma \autoref{lm:same_with_all_base} there exists an $(m,q)$-composite code $\mathcal{C}''\in\codeconbase$ such that
    \[\favg(\mathcal{C}'',n,\Dmld)\geq\favg(\mathcal{C}',n,\Dmld).
    \] 
    Therefore, there exists an $(m,q)$-composite code $\mathcal{C}''\in\codeconbase$ such that
     \[\favg(\mathcal{C}'',n,\Dmld)>\favg(\mathcal{C},n,\Dmld),
    \] 
    which results with a contradiction to the fact that $\mathcal{C}$ was optimal in $\codeconbase$. Hence $\mathcal{C}\in\optmin^n(q,m)\neq\phi$.
    
    $\Leftarrow$ Assume $\optmin^n(q,m)\neq\phi$, and let $\mathcal{C}\in\optmin^n(q,m)$. From Lemma \autoref{lm:same_with_all_base} there exists an $(m,q)$-composite code $\mathcal{C}'\in\codeconbase$ such that
    \[\favg(\mathcal{C}',n,\Dmld)\geq\favg(\mathcal{C},n,\Dmld).
    \] 
    Since $\codeconbase\subseteq\mathcal{C}_q^m$, then for every $\mathcal{C}''\in\codeconbase$ it holds that    
    \begin{align*}
    \favg(\mathcal{C}',n,\Dmld)\geq&\favg(\mathcal{C},n,\Dmld)\\
    \geq&\favg(\mathcal{C}'',n,\Dmld),
    \end{align*}
    because $\mathcal{C}$ is maximal in $\mathcal{C}_q^m$. Hence, by definition 
    \[
    \mathcal{C}'\in\arg\max_{\mathcal{C}\in\codeconbase}\favg(\mathcal{C},n,\Dmld)\neq\phi.
    \]
    \end{proof}

The next theorem finds an optimal code for the case where $m=q+1$.
Denote $\coptq\triangleq E_q\cup\{(\frac{1}{q},\dots,\frac{1}{q})\}$.
\begin{theorem}
For any $q,n\in\mathbb{N}$  it holds that 
\begin{enumerate}
    \item $\coptq\in\optavg^n(q+1,q)$,
    \item $\coptq\in\optmin^n(q+1,q)$,
    \item $\optavgprob^n(q+1,q)=1-\frac{1}{q^{n-1}(q+1)}$,
    \item $\optminprob^n(q+1,q)=1-\frac{1}{q^{n-1}}$.
\end{enumerate}
\end{theorem}
\begin{proof}

From Claim \autoref{cl:search_only_with_base} we know that it is sufficient to find an optimal $(q+1,q)$-composite code over $\mathcal{BC}_q^{q+1}$. Let $\mathcal{C}=\{E_q\cup\{\boldsymbol{c}_{k+1}\}\}$ be a general $(q+1,q)$-composite code in $\mathcal{BC}_q^{q+1}$, denote $\boldsymbol{c}_{k+1}=(c_1,\dots,c_q)$. Using Lemma \autoref{lm:perfect_decode_base_composite} for every $i\in[q]$ it holds that $$\psucc(\mathcal{C},n,\Dmld,\bfe_{q,i})=1,\bfe_{q,i}\in\drmld(\bfe_{q,i}).$$
    For every $\bfth\in\Omega_n^q\setminus{E_q}$ it holds for every $i\in[q]$ that $\mathbf{P}[\theta(\mathbf{X}^{\bfe_{q,i}}_{n})= \bfth]=0.$
    Hence, $\psucc(\mathcal{C},n,\Dmld,\boldsymbol{c}_{k+1})$ equals to
    \begin{align*}
    &\mathbf{P}[(\Theta(\mathbf{X}^{\boldsymbol{c}_{k+1}}_{n}))\in\Omega_n^q\setminus E_q]=1-\mathbf{P}[(\Theta(\mathbf{X}^{\boldsymbol{c}_{k+1}}_{n}))\in E_q]\\
    =&1-\sum_{i=1}^{q}\mathbf{P}[(\Theta(\mathbf{X}^{\boldsymbol{c}_{k+1}}_{n})) = \bfe_{q,i}]=1-\sum_{i=1}^{q}{(c_i)^n}=\epsilon_{\boldsymbol{c}}.
    \end{align*}
    Now we have
    \[
    \arg\max_{\mathcal{C}\in\mathcal{BC}_q^m}\favg(\mathcal{C},n,\Dmld)
    =\{E_q\cup\{\arg\max_{\boldsymbol{c}\in\Sigmaexp}{\frac{q+\epsilon_{\boldsymbol{c}}}{q+1}}\}\},
    \]
    \[\arg\max_{\mathcal{C}\in\mathcal{BC}_q^m}\fmin(\mathcal{C},n,\Dmld)
    =\{E_q\cup\{\arg\max_{\boldsymbol{c}\in\Sigmaexp}{\epsilon_{\boldsymbol{c}}}\}\}.
    \]
    Since $q\geq0$ we have that $${\arg\max_{\boldsymbol{c}\in\Sigmaexp}{\frac{q+\epsilon_{\boldsymbol{c}}}{q+1}}=\arg\max_{\boldsymbol{c}\in\Sigmaexp}{\epsilon_{\boldsymbol{c}}}}.$$
    We are left with the following optimization problem:
    $$
    \max_{\boldsymbol{c}\in\mathbb{R}^{q}}\big\{1-\sum_{i=1}^{q}{(c_i)^n}\big\} \mbox{ s.t. } \sum_{i=1}^{q}{c_i}=1.$$
    Using lagrange multiplier we have 
    $$
    \mathcal{L}(\boldsymbol{c}, \lambda)=1-\sum_{i=1}^{q}{(c_i)^n}+\lambda(\sum_{i=1}^q{c_i}-1).$$
    Thus, we get that $$\frac{\partial \mathcal{L}}{\partial c_i}=-{n(c_i)^{n-1}}+\lambda=0,$$ hence, $$\frac{\partial \mathcal{L}}{\partial \lambda}=\sum_{i=1}^q{c_i}-1=0.$$
    Therefore, $$c_1=\dots=c_q,$$ hence, $$\sum_{i=1}^q{c_i}=1.$$
    As a result, $$c_1=\dots=c_q=\frac{1}{q}.$$
    Hence we prove $(1)$ and $(2)$, i.e.,  $$\coptq\in\optavg^n(q,q+1),\coptq\in\optmin^n(q,q+1).$$
    One can calculate $(3)$ and $(4)$, i.e.,  $$\favg(\coptq,n,\Dmld)=1-\frac{1}{q^{n-1}(q+1)},$$ and $$\fmin(\coptq,n,\Dmld)=1-{(\frac{1}{q})}^{n-1}.$$
\end{proof}

The following theorem studies the case where $m$ is maximal.
First, we will prove the following lemma 
\begin{lemma}\label{lm:max_in_self}
    For any composite symbol $\bfrh'\in\Sigmaexp$, it holds for any $n\in\mathbb{N}$ that
    \[
    \arg\max_{\bfrh\in\Sigmaexp}\mathbf{P}[\theta(\mathbf{X}^{\bfrh}_{n})= \bfrh']=\bfrh'.
    \]
\end{lemma}
\begin{proof}
    Denote $\bfrh=(\rho_1,\dots,\rho_q)$, $\bfrh'=(\rho_1',\dots,\rho_q')$.
    \[
        \mathbf{P}[\theta(\mathbf{X}^{\bfrh}_{n})=\bfrh']=\binom{n}{\rho_1',\dots,\rho_q'n}\prod_{j=1}^q(\rho_j)^{\rho_j'n}
    \]
    Since the binomial coefficient is constant and $n\geq1$ we can optimize only
    \[
    \Pi_{j=1}^q(\rho_j)^{\rho_j'}
    \]
    We are left with the following optimisation problem:
    \[
    \mbox{max}_{\bfrh\in\mathbb{R}^{q}} \Pi_{j=1}^q(\rho_j)^{\rho_j'} \mbox{ s.t. } \sum_{i=1}^{q}{\rho_i}=1. 
    \]
    Using lagrange multiplier we have
    \[
    \mathcal{L}(\bfrh, \lambda)=\Pi_{j=1}^q(\rho_j)^{\rho_j'}+\lambda(\sum_{i=1}^q{\rho_i}-1).
    \]
    Thus, we get that
    \begin{gather*}
    \frac{\partial \mathcal{L}}{\partial \rho_i}=\rho_i'(\rho_i)^{\rho_i'-1}\Pi_{j=1,j\neq i}^q(\rho_j)^{\rho_j'}+\lambda=0 \\
    \frac{\partial \mathcal{L}}{\partial \lambda}=\sum_{i=1}^q{\rho_i}-1=0
    \end{gather*}
    Denote $\alpha(\bfrh)=\Pi_{j=1}^q(\rho_j)^{\rho_j'}$, it holds that
    \[
    \frac{\rho_i'}{\rho_i}=-\frac{\lambda}{\alpha(\bfrh)},
    \]
    hence,
    \[
    \rho_i=-\frac{\alpha(\rho)}{\lambda}\rho_i'.
    \]
    Since
    \[
    \sum_{i=1}^q{\rho_i'}=1=\sum_{i=1}^q{\rho_i'},
    \]
    we get that
    \[
    -\frac{\lambda}{\alpha(\bfrh)}=1,
    \]
    therefore,
    \[
    \rho_i=\rho_i'.
    \]
    Thus, we get that
    \[
    \arg\max_{\bfrh\in\Sigmaexp}\mathbf{P}[\theta(\mathbf{X}^{\bfrh}_{n})= \bfrh']=\bfrh'.
    \]

\end{proof}

The previous lemma proves that if the MLD receives a certain distribution and the distribution itself is a symbol then it always chooses this symbol. Hence we can conclude the following corollary.
\begin{corollary}\label{cor:max_in_self}
    For any $(m,q)$-composite code $\mathcal{C}$ and $\boldsymbol{c}\in\mathcal{C}$ it holds that
    \[
    \boldsymbol{c}\in\drmld(\boldsymbol{c}).
    \]
\end{corollary}

We denote by $\ccnk$ the number of integer solutions of the equation $x_1+\dots+x_q=n \mbox{ s.t. } x_i\geq0$.
For any $\bfrh\in\Sigmaexp$ denote 
\[
\beta_n^{\bfrh}\triangleq\binom{n}{\rho_1 n,\dots ,\rho_q n}\prod_{j=1}^q(\rho_j)^{\rho_jn}
\]
\begin{theorem}
For any $q,n\in\mathbb{N}$ it holds that 
\begin{enumerate}
    \item $\Omega_n^q\in\optavg^n(\ccnk,q)$,
    \item $\optavgprob(\ccnk,q)=\frac{1}{\ccnk}\sum_{\bfrh\in\Omega_n^q}{\beta_n^{\bfrh}}$,
    \item $\Omega_n^q\in\optmin^n(\ccnk,q)$,
    \item $\optminprob(\ccnk,q)=\min_{\bfrh\in\Omega_n^q}{\beta_n^{\bfrh}}$.
\end{enumerate}
\end{theorem}

\begin{proof}
First, one can calculate that $$|\Omega_n^q|=\ccnk.$$
Let $\mathcal{C}$ be a $(\ccnk,q)$-composite code.
Using Lemma \autoref{lm:split_avg_to_sum} we have that
\[
\favg(\mathcal{C},n,\Dmld)=\frac{1}{\ccnk}\sum_{\bfth\in\Omega_n^q}\mathbf{P}[\Theta(\mathbf{X}^{\Dmld{\theta})}_n)=\bfth]
\]
Using Lemma \autoref{lm:max_in_self}, we now have that
\[
\leq\frac{1}{\ccnk}\sum_{\bfth\in\Omega_n^q}\mathbf{P}[\Theta(\mathbf{X}^{\bfth}_n)=\bfth].
\]
Using \autoref{cor:max_in_self} for the $(\ccnk,q)$-composite code $\Omega_n^q$ results with
\[
\favg(\Omega_n^q,n,\Dmld)=\frac{1}{\ccnk}\sum_{\bfth\in\Omega_n^q}\mathbf{P}[\Theta(\mathbf{X}^{\bfth}_n)=\bfth].
\]
Therefore, for any $(\ccnk,q)$-composite code $\mathcal{C}$, it holds that
\[
\favg(\mathcal{C},n,\Dmld)\leq\favg(\Omega_n^q,n,\Dmld).
\]
Hence, by definition we prove $(1)$, i.e.,
\[
\Omega_n^q\in\optavg^n(\ccnk,q)
\]
In order to prove $(2)$ one can note that,
\begin{align*}
    \favg(\Omega_n^q,n,\Dmld)=&\frac{1}{\ccnk}\sum_{\bfth\in\Omega_n^q}\mathbf{P}[\Theta(\mathbf{X}^{\bfth}_n)=\bfth]
    =\frac{1}{\ccnk}\sum_{\bfth\in\Omega_n^q}{\beta_n^{\bfth}}.
\end{align*}
Furthermore, Let $\mathcal{C}$ be an $(\ccnk,q)$-composite code. Since $|\Omega_n^q|=\ccnk$ it holds that for every $\boldsymbol{c}\in\mathcal{C}$, $\drmld(\boldsymbol{c})\neq\phi$ if and only if for every $\boldsymbol{c}\in\mathcal{C}$ $|\drmld(\boldsymbol{c})|=1$. If there exists $\boldsymbol{c}\in\mathcal{C}$ such that $\drmld(\boldsymbol{c})=\phi$ then for every $n\in\mathbb{N}$
\[
\psucc(\mathcal{C},n,\Dmld,\boldsymbol{c})=0.
\]
Hence,
 \[\fmin(\mathcal{C},n,\Dmld)=\min_{\boldsymbol{c}\in\mathcal{C}}\psucc(\mathcal{C},n,\Dmld,\boldsymbol{c})=0.
\]
Therefore, we can assume that for every $\boldsymbol{c}\in\mathcal{C}$ $|\drmld(\boldsymbol{c})|=1$, and denote each $\boldsymbol{c}_{\bfth}\in\mathcal{C}$ by its DR i.e., for every $\bfth\in\Omega_n^q$, $\boldsymbol{c}_{\bfth}\in\mathcal{C}$ and
\[
\drmld(\boldsymbol{c}_{\bfth})=\{\bfth\}.
\]
Therefore,
\[
\fmin(\Omega_n^q,n,\Dmld)=\min_{\bfth\in\Omega_n^q}\mathbf{P}[\Theta(\mathbf{X}^{\boldsymbol{c}_{\bfth}}_n)=\bfth].
\]
For the $(\ccnk,q)$-composite code $\Omega_n^q$ we have
\[
\fmin(\Omega_n^q,n,\Dmld)=\min_{\bfth\in\Omega_n^q}\mathbf{P}[\Theta(\mathbf{X}^{\bfth}_n)=\bfth].
\]
Using Lemma \autoref{lm:max_in_self} we now have that for every $\bfth\in\Omega_n^q$
\[
\mathbf{P}[\Theta(\mathbf{X}^{\boldsymbol{c}_{\bfth}}_n)=\bfth]\leq
\mathbf{P}[\Theta(\mathbf{X}^{\bfth}_n)=\bfth]
\]
Hence, 
\[
\fmin(\Omega_n^q,n,\Dmld)\leq\fmin(\Omega_n^q,n,\Dmld).
\]
Therefore, for any $(\ccnk,q)$-composite code $\mathcal{C}$, it holds that
\[
\fmin(\mathcal{C},n,\Dmld)\leq\fmin(\Omega_n^q,n,\Dmld).
\]
Hence, by definition we prove $(3)$, i.e.,
\[
\Omega_n^q\in\optmin^n(\ccnk,q)
\]
In order to prove $(4)$ one can note that,
\begin{align*}
    \fmin(\Omega_n^q,n,\Dmld)=&\min_{\bfth\in\Omega_n^q}\mathbf{P}[\Theta(\mathbf{X}^{\bfth}_n)=\bfth]
    =\min_{\bfth\in\Omega_n^q}{\beta_n^{\bfth}}.
\end{align*}

\end{proof}

\begin{theorem}
     For any $q\in\mathbb{N}$ and $n=rq$ for $r\in\mathbb{N}$ 
     \[\optminprob\left(\ccnk,q\right)=\binom{n}{r, \dots, r}(\frac{1}{q})^{n}.\]
\end{theorem}
\begin{proof}
Denote $\bfrh_r=(\frac{r}{n},\dots,\frac{r}{n})$. It holds that 
\[
\beta_n^{\bfrh_r}=\binom{n}{r, \dots, r}(\frac{1}{q})^{n}.
\]
We will show that for every $\bfrh\in\Omega_n^q$ such that $\bfrh\neq\bfrh_r$ there exists $\bfrh'\in\Omega_n^q$ such that
\[
\beta_n^{\bfrh'}\leq\beta_n^{\bfrh}
\]
and $||\bfrh_r-\bfrh'||_2<||\bfrh_r-\bfrh||_2$. As a result we will have that \[
\min_{\bfth\in\Omega_n^q}{\beta_n^{\bfth}}=\beta_n^{\bfrh_r}.
\]
    Let $\bfrh\in\Omega_n^q$ be such that $\bfrh\neq\bfrh_r$. Denote $\bfrh=(\frac{i_1}{n},\frac{i_2}{n},\dots,\frac{i_q}{n})$ such that $i_1,\dots,i_q\in\mathbb{N}$ and $\sum_{j=1}^q{i_j}=n$. Since $\bfrh\neq\bfrh_r$ there exists an index $j_1$ such that $i_{j_1}\neq r$. If $i_{j_1}<r$, then there exists an index $j_2$ such that $i_{j_1}>r$ using the peigon hole principle. Similarly if $i_{j_1}>r$, there exist an index $j_2$ such that $i_{j_1}<r$. W.l.o.g we assume that $j_1=1,j_2=2$ and $i_1<r, i_2>r$.
Denote $\bfrh'=(\frac{i_1+1}{n},\frac{i_2-1}{n},\dots,\frac{i_q}{n})$, it holds that $\bfrh'\in\Omega_n^q$ and $||\bfrh_r-\bfrh'||_2=||\bfrh_r-\bfrh||_2-2$. Furthermore 
\[
\frac{\beta_n^{\bfrh'}}{\beta_n^{\bfrh}}=\frac{\binom{n}{i_1+1,i_2-1, \dots, r}(\frac{i_1+1}{n})^{i_1+1}(\frac{i_2-1}{n})^{i_2-1}\prod_{j\neq1,2}{\rho_j}^{n\rho_j}}{\binom{n}{i_1,i_2, \dots, r}(\frac{i_1}{n})^{i_1}(\frac{i_2}{n})^{i_2}\prod_{j\neq1,2}{\rho'_j}^{n\rho_j}}.
\]
Since $\rho_j=\rho'_j$ for $j\notin{1,2}$ we have that
\begin{align*}
    \frac{\beta_n^{\bfrh'}}{\beta_n^{\bfrh}}=&\frac{\binom{n}{i_1+1,i_2-1, \dots, r}(\frac{i_1+1}{n})^{i_1+1}(\frac{i_2-1}{n})^{i_2-1}}{\binom{n}{i_1,i_2, \dots, r}(\frac{i_1}{n})^{i_1}(\frac{i_2}{n})^{i_2}}
    =\frac{\frac{n!}{(i_1+1)!(i_2-1)!\dots i_q!}(\frac{i_1+1}{n})^{i_1+1}(\frac{i_2-1}{n})^{i_2-1}}{\frac{n!}{i_1!i_2!\dots i_q!}(\frac{i_1}{n})^{i_1}(\frac{i_2}{n})^{i_2}}\\
    =&\frac{i_1!i_2!}{(i_1+1)!(i_2-1)!}\frac{i_1+1}{n}\frac{n}{i_2}(\frac{i_1+1}{i_1})^{i_1}(\frac{i_2-1}{i_2})^{i_2-1}
    =(\frac{i_1+1}{i_1})^{i_1}(\frac{i_2-1}{i_2})^{i_2-1}.
\end{align*}
It holds that $i_1<r<i_2$, therefore, $i_1+1\leq i_2$, hence,
\begin{align*}
    \frac{\beta_n^{\bfrh'}}{\beta_n^{\bfrh}}=&(\frac{i_1+1}{i_1})^{i_1}(\frac{i_2-1}{i_2})^{i_2-1}
    \leq(\frac{i_2}{i_1})^{i_2-1}(\frac{i_2-1}{i_2})^{i_2-1}
    =(\frac{i_2-1}{i_1})^{i_2-1} \leq 1.
\end{align*}
Therefore, $\beta_n^{\bfrh'}\leq\beta_n^{\bfrh}$.
\end{proof}

\begin{theorem}
For any $q,n\in\mathbb{N}$ it holds for any $m>\ccnk$ that 
\begin{enumerate}
    \item $\optavgprob^n(m,q)=\frac{1}{m}\sum_{\bfrh\in\Omega_n^q}{\beta_n^{\bfrh}}$,
    \item $\optminprob^n(m,q)=0$.
\end{enumerate}
\end{theorem}
\begin{proof}
$(1)$
Let $\mathcal{C}$ be an $(m,q)$-composite code for $m>\ccnk$.
Using Lemma \autoref{lm:split_avg_to_sum} we have that
\[
\favg(\mathcal{C},n,\Dmld)=\frac{1}{m}\sum_{\bfth\in\Omega_n^q}\mathbf{P}[\Theta(\mathbf{X}^{\Dmld{\theta})}_n)=\bfth].
\]
Using Lemma \autoref{lm:max_in_self} we now have that
\[
\leq\frac{1}{m}\sum_{\bfth\in\Omega_n^q}\mathbf{P}[\Theta(\mathbf{X}^{\bfth}_n)=\bfth]=\frac{1}{m}\sum_{\bfth\in\Omega_n^q}{\beta_n^{\bfth}}.
\]
Hence,
\[
\optavgprob(m,q)\leq\frac{1}{m}\sum_{\bfrh\in\Omega_n^q}{\beta_n^{\bfrh}}.
\]
Using \autoref{cor:max_in_self} for any $(m,q)$-composite code that contains $\Omega_n^q$ results with
\begin{align*}
    \favg(\Omega_n^q,n,\Dmld)=&\frac{1}{m}\sum_{\bfth\in\Omega_n^q}\mathbf{P}[\Theta(\mathbf{X}^{\bfth}_n)=\bfth]\\
    =&\frac{1}{m}\sum_{\bfth\in\Omega_n^q}{\beta_n^{\bfth}}.
\end{align*}
Therefore,
\[
\optavgprob(m,q)=\frac{1}{m}\sum_{\bfrh\in\Omega_n^q}{\beta_n^{\bfrh}}.
\]

$(2)$
    Let $\mathcal{C}$ be an $(m,q)$-composite code for $m>\ccnk$. Since $|\Omega_n^q|=\ccnk$ it holds that there exists $\boldsymbol{c}\in\mathcal{C}$ such that
    \[
    \drmld(\boldsymbol{c})=\phi.
    \]
    Hence,
    \[
    \fmin(\mathcal{C},n,\Dmld)\leq\mathbf{P}[\Theta(\mathbf{X}^{\boldsymbol{c}}_n)=\boldsymbol{c}]=0.
    \]
    Since
    \[
     \fmin(\mathcal{C},n,\Dmld)\geq0,
    \]
    it holds that
    \[
    \fmin(\mathcal{C},n,\Dmld)=0.
    \]
    Since $\mathcal{C}$ was arbitrary we have that 
    \[
    \optminprob^n(m,q)=0.
    \]
\end{proof}

\section{Results for the Binary Alphabet}\label{sec:binary}
We call an $(m,2)$-composite code $\mathcal{C}$ a \emph{binary code}. Since every codeword $\boldsymbol{c}\in\mathcal{C}$ is a distribution $(\alpha, 1-\alpha)$, each codeword will be denoted by its first value in the distribution.
For an $(m,2)$-composite code $\mathcal{C}$, we always order the codewords in $\mathcal{C}$ by their value and denote $\mathcal{C}=\{\bfc_1,\dots,\bfc_m\}$, i.e., for every $i<j$ $\bfc_i=(c_i,1-c_i), \bfc_j=(c_j,1-c_j)$ and it holds that $c_i<c_j$. For a composite symbol $\bfrh\in\compbin$ the composite symbol  $\overline{\bfrh}$ is its \emph{symmetric composite symbol} if $\overline{\rho}=1-\rho$.
For a set of composite symbols $P\subseteq\compbin$ the set $\overline{P}$ is the \emph{symmetric set} of $P$ if it contains all its symmetric symbols, i.e., $\overline{P}=\{\overline{\bfrh}:\bfrh\in P\}$. Let $\cC=\{\bfc_1,\dots,\bfc_m\}$ be a binary composite code, denote by $\overline{\cC}$ the \emph{symmetric code} of $\cC$ that includes all the symmetric symbols of symbols in $\cC$, i.e., $\overline{\cC}=\{\overline{\bfc}_1,\dots,\overline{\bfc}_m\}$. Furthermore we denote by $\Dmldadj$, $\drmldadj$ the MLD and the DR of $\overline{\cC}$, respectively. A binary composite code $\mathcal{C}$ is called a \emph{symmetric composite code} if $\cC=\overline{\cC}$. Denote by $\symcode$ the set of all symmetric $(m,2)$-composite codes and by $\symcodeconbase$ the set of all symmetric $(m,2)$-composite codes that contain $E_2$.

The following claim justifies narrowing the search for optimal binary codes only to symmetric codes that contain the base composite alphabet.
First we will prove the following lemma, its full proof is in Appendix~\ref{app:B}. 
\begin{restatable}{lemma}{sameminwithsymLm}\label{lm:same_min_with_sym}
    For any code $\cC\in\codebinconbase^{2m}$ there exists a code $\cC\in\symevencodeconbase$ such that
    \[\fmin(\mathcal{C}',n,\Dmld)\geq\fmin(\mathcal{C},n,\Dmld).
    \]
\end{restatable}

\begin{claim}\label{cl:search_only_symetric}
For any $m,n\in\N$ it holds that $\optmin^n(2,2m)=\phi$ if and only if $\arg\max_{\cC\in\symevencodeconbase}\fmin(\mathcal{C},n,\Dmld)=\phi$.
\end{claim}
\begin{proof}
Using Claim \autoref{cl:search_only_with_base} we have that
$\optmin^n(2,2m)=\phi\mbox{ if and only if  }\arg\max_{\mathcal{C}\in\codebinconbase^{2m}}\fmin(\mathcal{C},n,\Dmld)=\phi$.

Therefore, it is sufficient to prove that
$\arg\max_{\cC\in\symevencodeconbase}\fmin(\mathcal{C},n,\Dmld)=\phi$ if and only if  $\arg\max_{\mathcal{C}\in\codebinconbase^{2m}}\fmin(\mathcal{C},n,\Dmld)=\phi$.

$\Rightarrow$ Assume $\arg\max_{\cC\in\symevencodeconbase}\fmin(\mathcal{C},n,\Dmld)\neq\phi$, and let $\mathcal{C}\in\arg\max_{\cC\in\symevencodeconbase}\fmin(\mathcal{C},n,\Dmld)$. If $\mathcal{C}\notin\arg\max_{\mathcal{C}\in\codebinconbase^{2m}}\fmin(\mathcal{C},n,\Dmld)$ then there exist a binary code $\mathcal{C}'\in\codebinconbase^{2m}$ such that
    \[\fmin(\mathcal{C}',n,\Dmld)>\fmin(\mathcal{C},n,\Dmld).
    \]
    From Lemma \autoref{lm:same_min_with_sym} there exist a symmetric binary code $\mathcal{C}''\in\symevencodeconbase$ such that
    \[\fmin(\mathcal{C}'',n,\Dmld)\geq\fmin(\mathcal{C}',n,\Dmld).
    \] 
    Therefore, there exist a symmetric binary code $\mathcal{C}''\in\symevencodeconbase$ such that
     \[\fmin(\mathcal{C}'',n,\Dmld)\geq\fmin(\mathcal{C},n,\Dmld),
    \] 
    thus, we get a contradiction to the fact that $\mathcal{C}$ is maximal in $\symevencodeconbase$. Hence $\mathcal{C}\in\codebinconbase^{2m}\neq\phi$.
    
$\Leftarrow$ Assume $\arg\max_{\mathcal{C}\in\codebinconbase^{2m}}\fmin(\mathcal{C},n,\Dmld)\neq\phi$, and let $\mathcal{C}\in\arg\max_{\mathcal{C}\in\codebinconbase^{2m}}\fmin(\mathcal{C},n,\Dmld)$. From Lemma \autoref{lm:same_min_with_sym} there exist a symmetric binary code $\mathcal{C}'\in\symevencodeconbase$ such that
    \[\fmin(\mathcal{C}',n,\Dmld)\geq\fmin(\mathcal{C},n,\Dmld).
    \] 
    Since $\symevencodeconbase\subseteq\codebinconbase^{2m}$, then for every $\mathcal{C}''\in\symevencodeconbase$ it holds that
    \begin{align*}
    \fmin(\mathcal{C}',n,\Dmld)\geq&\fmin(\mathcal{C},n,\Dmld)\\
    \geq&\fmin(\mathcal{C}'',n,\Dmld).
    \end{align*}
    because $\mathcal{C}$ is maximal in $\codebinconbase^{2m}$. Hence by definition 
    \[
    \mathcal{C}'\in\arg\max_{\cC\in\symevencodeconbase}\fmin(\mathcal{C},n,\Dmld)\neq\phi.
    \]
    \end{proof}

Now we find an optimal alphabet of size $4$ for every $n\in\N$.

\begin{theorem}
For any $m\in\N$ denote 
$$\beta_n=
    \begin{cases}
      \sqrt[m]{\frac{m+1}{2m+1}\binom{2m+1}{m}}, & n=2m+1,\\
      \sqrt[m]{\frac{m+1}{2m+1}\binom{2m}{m-1}}, & n=2m.
    \end{cases}$$
For any $n\in\mathbb{N}$ denote $\alpha(n)=\frac{1}{1+\beta_n}$. Then, it holds that $$\{0,\alpha(n),1-\alpha(n),1\}\in\optmin^n(4,2).$$
\end{theorem}
\begin{proof}
Let $n\in\N$ be an odd number $n=2m+1$. Using Claim \autoref{cl:search_only_symetric}, if $\optmin^n(2,4)\neq\phi$ then $\arg\max_{\cC\in\symcodeconbasefour}\fmin(\mathcal{C},n,\Dmld)\neq\phi$ so it is sufficient to find an optimal symmetric composite code that contains $E_2$, i.e.,
    $\arg\max_{x\in(0,\frac{1}{2})}\fmin(\{0,x,1-x,1\},n,\Dmld).$
    Using Lemma {\autoref{lm:perfect_decode_base_composite}}, every base composite symbol has perfect decoding probability. The symbols with probabilities $x,1-x$ are symmetric in a symmetric code, since $\frac{1}{2}$ cannot be seen since $n$ is even we have that they are equal. Therefore, it holds that 
    $$\fmin(\{0,x,1-x,1\},n,\Dmld)=\psucc(\bfx).$$
    Furthermore, $$\psucc(\bfx)=\sum_{i=1}^{m}{\mathbf{P}[\Theta(\mathbf{X}^{\bfx}_n)=(\frac{i}{n},\frac{n-i}{n})]}.$$ Let $X$ be a random variable such that $X\sim\mbox{Bin}(p)$.
    It is known that $$\mathbf{P}[X\leq k]=(n-k)\binom{n}{k}\int_{0}^{1-x}{t^{n-k-1}(1-t)^k}dt.$$
    For $n=2m+1$ we set $p=x,k=m$, thus we get $$\psucc(\bfx)=(m+1)\binom{2m+1}{m}\int_{0}^{1-x}{t^{m}(1-t)^m}dt-(1-x)^n.$$
    Denote $$\psi(x)=(m+1)\binom{2m+1}{m}\int_{0}^{1-x}{t^{m}(1-t)^m}dt-(1-x)^n.$$
    We have that
    \begin{align*}
    &\psi'(x)=(m+1)\binom{2m+1}{m}\left(\int_{0}^{1-x}{t^{m}(1-t)^m}dt\right)'+n(1-x)^{n-1}\\
    &=-(m+1)\binom{2m+1}{m}{(1-x)^{m}x^m}+(2m+1)(1-x)^{2m}.
    \end{align*}
    By setting $\psi'(x)=0$ we get that $$(m+1)\binom{2m+1}{m}{(1-x)^{m}x^m}=(2m+1)(1-x)^{2m}.$$
    Since $x\neq0$ and we get $$\frac{m+1}{2m+1}\binom{2m+1}{m}=({\frac{1-x}{x}})^{m}.$$
    Now we have $\frac{1-x}{x}=\beta_n$.
    Therefore, $$x=\frac{1}{1+\beta_n}.$$
    For $n=2m$ we set $p=x,k=m$, thus we get that $\psucc(\bfx)$ equals to 
    $$m\binom{2m}{m}\int_{0}^{1-x}{t^{m}(1-t)^m}dt-(1-x)^n-\binom{2m}{m}(\frac{1}{2})^{2m}$$
    Denote $$\psi(x)=m\binom{2m}{m}\int_{0}^{1-x}{t^{m}(1-t)^m}dt-(1-x)^n-\binom{2m}{m}(\frac{1}{2})^{n}.$$
    We have that
    \begin{align*}
    &\psi'(x)=m\binom{2m}{m}(\int_{0}^{1-x}{t^{m}(1-t)^m}dt)'+n(1-x)^{n-1}
    =-m\binom{2m}{m}{(1-x)^{m}x^m}+(2m)(1-x)^{2m-1}.
    \end{align*}
    By setting $\psi'(x)=0$ we get that $$(m+1)\binom{2m}{m-1}{(1-x)^{m}x^m}=(2m)(1-x)^{2m-1}.$$
    Since $x\neq0$ and we get $$\frac{m+1}{2m+1}\binom{2m}{m-1}=({\frac{1-x}{x}})^{m}.$$
    Now we have $\frac{1-x}{x}=\beta_n$.
    Therefore, $$x=\frac{1}{1+\beta_n}.$$
\end{proof}

In conclusion we found the optimal composite alphabet of size $4$ over the binary alphabet for any size of transmission $n$. For $n\rightarrow\infty$ it holds that
\[
\beta_n\rightarrow 4.
\]
Therefore, for $n\rightarrow\infty$ the optimal alphabet converges to $(0,\frac{1}{5},\frac{4}{5}, 1)$.

\newpage

\begin{appendices}

\section{}\label{app:A}

\lmaperfectbaseLm*
\begin{proof}
    from Lemma \autoref{lm:supp_subset_orig} we know that
    \[
    \imn(\bfe_{q,i})\subseteq\{\bfe_{q,i}\},
    \]
    since the $n$-transmission of $\bfe_{q,i}$ always results with the observed distribution of $\bfe_{q,i}$ we have that
    \[
    \imn(\bfe_{q,i})=\{\bfe_{q,i}\}.
    \]
    Using \autoref{cor:max_in_self} we get that
    \[
    \bfe_{q,i}\in\drmld(\bfe_{q,i}),
    \]
    and hence $\psucc(\mathcal{C},n,\Dmld,\bfe_{q,i})$ is equal to
    \[
    \mathbf{P}[(\Theta(\mathbf{X}^{\bfe_{q,i}}_{n}))\in\imn(\bfe_{q,i})\cap\drmld(\bfe_{q,i})
    =\mathbf{P}[(\Theta(\mathbf{X}^{\bfe_{q,i}}_{n}))\in\imn(\bfe_{q,i}))]=1.
    \]
\end{proof}

\begin{lemma}\label{lmabaseadd}
    For any $(m,q)$-composite code $\mathcal{C}$, a composite symbol $\boldsymbol{c'}\in\mathcal{C}$ and $e_{q,i}\in\drmld(\boldsymbol{c'})$ denote by $\mathcal{C}'=(\mathcal{C}\setminus\{\boldsymbol{c'}\})\cup\{\bfe_{q,i}\}$. It holds that
    \begin{itemize}
        \item $\fmin(\mathcal{C'},n,\Dmld)\geq\fmin(\mathcal{C},n,\Dmld)$
        \item $\favg(\mathcal{C'},n,\Dmld)\geq\favg(\mathcal{C},n,\Dmld)$
    \end{itemize}
\end{lemma}
\begin{proof}
    We will prove that for every $\boldsymbol{c}\in\mathcal{C}\cap\mathcal{C}'$ it holds for any $n\in\mathbb{N}$ that
    \[
    \psucc(\mathcal{C}',n,\Dmld,\boldsymbol{c})\geq\psucc(\mathcal{C},n,\Dmld,\boldsymbol{c}).
    \]
    From Lemma \autoref{lm:perfect_decode_base_composite} for any $n\in\mathbb{N}$
    \[
    \psucc(\mathcal{C}',n,\Dmld,\bfe_{q,i})=1,
    \]
    and from the fact that $\psucc(\mathcal{C},n,\Dmld,\boldsymbol{c})\leq1$ we conclude that
    \[
    \fmin(\mathcal{C'},n,\Dmld)\geq\fmin(\mathcal{C},n,\Dmld),
    \favg(\mathcal{C'},n,\Dmld)\geq\fmin(\mathcal{C},n,\Dmld).
    \]
    We denote $\Dmld, \drmld$ for the code $\mathcal{C}$ and $\Dmld', \drmld'$ for the code $\mathcal{C}'$. Let $\boldsymbol{c}''\in\mathcal{C}\cap\mathcal{C}'$ we will prove that for every $n\in\mathbb{N}$
    \[
    \drmld(\boldsymbol{c}'')\subseteq\drmld'(\boldsymbol{c}'').
    \]
    We know that
    \[
    \Dmld(\bfe_{q,i}, n)= \mbox{argmax}_{\boldsymbol{c}\in\mathcal{C}}{\mathbf{P}[(\Theta(\mathbf{X}^{\boldsymbol{c}}_n))=\bfe_{q,i}]}=\boldsymbol{c}',
    \]
    and therefore $\bfe_{q,i}\notin\drmld(\boldsymbol{c}'')$. Now for every $\bfth\in\drmld(\boldsymbol{c}'')$ we have $\bfth\neq\bfe_{q,i}$, therefore  $\bfth=(\theta_1,\dots,\theta_q)$ for $\theta_i<1$ and hence,
    \[
     \mathbf{P}[(\Theta(\mathbf{X}^{\bfe_{q,i}}_n))=\bfth]=0.
    \]
    Thus, we get that
    \[
    \boldsymbol{c}''= \mbox{argmax}_{\boldsymbol{c}\in\mathcal{C}'}\mathbf P[(\Theta(\mathbf{X}^{\boldsymbol{c}}_n))=\bfth],
    \]
    \[
    \drmld(\boldsymbol{c}'')\subseteq\drmld'(\boldsymbol{c}'').
    \]
    It hodls that
    \[
    \psucc(\mathcal{C},n,\Dmld,\boldsymbol{{c}''})=\mathbf{P}[(\Theta(\mathbf{X}^{\boldsymbol{c}''}_{n}))\in\imn(\boldsymbol{c}'')\cap\drmld(\boldsymbol{c}'')],
    \]
    \[
    \psucc(\mathcal{C}',n,\Dmld',\boldsymbol{{c}''})=\mathbf{P}[(\Theta(\mathbf{X}^{\boldsymbol{c}''}_{n}))\in\imn(\boldsymbol{c}'')\cap\drmld'(\boldsymbol{c}'')],
    \]
    therefore, will conclude that
    \[
    \psucc(\mathcal{C}',n,\Dmld,\boldsymbol{c})\geq\psucc(\mathcal{C},n,\Dmld,\boldsymbol{c}).
    \]
\end{proof}

From now on in this paper we denote by $[n]$ all the natural numbers between  $1$ and $n$, i.e.,
$
[n]=\{1,\dots,n\}.
$

\samewithallbaseLemma*
\begin{proof}
    Let $\mathcal{C}$ be an $(m,q)$-composite code  with $m\geq q$, we will prove that there exists a family of $(m,q)$-composite codes $\{\mathcal{C}_i\}_{i\in[m]}$ such that for every $i\in[m]$
    \begin{itemize}
        \item $\{\bfe_{q,j}\}_{j\in[i]}\subseteq\mathcal{C}_i$,
        \item $\favg(\mathcal{C}_i,n,\Dmld)\geq\favg(\mathcal{C},n,\Dmld)$,
        \item $\fmin(\mathcal{C}_i,n,\Dmld)\geq\fmin(\mathcal{C},n,\Dmld)$.
    \end{itemize}
    As a result, we will use $\mathcal{C}'=\mathcal{C}_m$ as an $(m,q)$-composite code such that
    \begin{itemize}
        \item $E_q\subseteq\mathcal{C}'$,
        \item $\favg(\mathcal{C'},n,\Dmld)\geq\favg(\mathcal{C},n,\Dmld)$,
        \item $\fmin(\mathcal{C'},n,\Dmld)\geq\fmin(\mathcal{C},n,\Dmld)$.
    \end{itemize}
    We will prove it using induction. 
    We denote $\Dmld, \drmld$ for the code $\mathcal{C}$ and $\Dmld^i, \drmld^i$ for the code $\mathcal{C}_i$.
    Base: $i=1$.
    Since
    \[\cup_{\boldsymbol{c}\in\mathcal{C}}\drmld^(\boldsymbol{c})=\Omega_n^q,
    \]
    there exists $\boldsymbol{c}\in\mathcal{C}$ such that 
    \[
    \bfe_{q,1}\in\drmld(\boldsymbol{c}).
    \]
    Using Lemma \autoref{lmabaseadd} we get that for $\mathcal{C}_1=(\mathcal{C}\setminus\{\boldsymbol{c}\})\cup\{\bfe_{q,1}\}$
    \begin{itemize}
        \item $\{\bfe_{q,1}\}\subseteq\mathcal{C}_1$,
        \item $\favg(\mathcal{C}_1,n,\Dmld)\geq\favg(\mathcal{C},n,\Dmld)$,
        \item $\fmin(\mathcal{C}_1,n,\Dmld)\geq\fmin(\mathcal{C},n,\Dmld)$.
    \end{itemize}
    Step: assume the claim is true for $i\leq k$, will prove for $i=k+1$.
    There exists a family of $(m,q)$-composite codes $\{\mathcal{C}_i\}_{i\in[k]}$ such that for every $i\in[k]$
    \begin{itemize}
        \item $\{\bfe_{q,j}\}_{j\in[i]}\subseteq\mathcal{C}_i$,
        \item $\favg(\mathcal{C}_i,n,\Dmld)\geq\favg(\mathcal{C},n,\Dmld)$,
        \item $\fmin(\mathcal{C}_i,n,\Dmld)\geq\fmin(\mathcal{C},n,\Dmld)$.
    \end{itemize}
    Since
    \[\cup_{\boldsymbol{c}\in\mathcal{C}_k}\drmld^{k}(\boldsymbol{c})=\Omega_n^q,
    \]
    there exists $\boldsymbol{c}\in\mathcal{C}_k$ such that 
    \[
    \boldsymbol{e_{q,n+1}}\in\drmld^k(\boldsymbol{c}).
    \]
    
    If $\boldsymbol{c}=\bfe_{q,i}$ for $i\leq n$, then since
    \[
    \psucc(\mathcal{C}_k,n,\Dmld^k,\bfe_{q,i})=\mathbf{P}[\Theta(\mathbf{X}^{\bfe_{q,i}}_{n})=\boldsymbol{e_{q,n+1}}]=0
    \]
    we can replace  $\boldsymbol{e_{q,n+1}}$ to be in any $\drmld^k(\boldsymbol{c})$ for any $\boldsymbol{c}\in\mathcal{C}_k$. Therefore, we choose $\boldsymbol{c}\in\mathcal{C}_k$ such that $\boldsymbol{c}\neq\bfe_{q,i}$ for $i\leq n$.

    Using Lemma \autoref{lmabaseadd} we get that for $\mathcal{C}_{k+1}=(\mathcal{C}_k\setminus\{\boldsymbol{c}\})\cup\{\boldsymbol{e_{q,k+1}}\}$
    \begin{itemize}
        \item $\{\bfe_{q,j}\}_{j\in[k+1]}\subseteq\mathcal{C}_{k+1}$,
        \item $\favg(\mathcal{C}_{k+1},n,\Dmld)\geq\favg(\mathcal{C},n,\Dmld)$,
        \item $\fmin(\mathcal{C}_{k+1},n,\Dmld)\geq\favg(\mathcal{C},n,\Dmld)$.
    \end{itemize}
    
\end{proof}

\section{}\label{app:B}

Let $\bfth$ be a composite symbol in $\compbin$. Let us present the following function
\[
\fth(x)=\mathbf{P}[\Theta(\mathbf{X}^{\boldsymbol{x}}_n)=\bfth]
\]
\begin{lemma}\label{lmaincdec}
For every $\bfth\in\compsinbase$ it holds that
\begin{itemize}
    \item The function $\fth(x)$ is differentiable in $(0,1)$.
    \item The function $\fth'(x)$ is positive in $(0,\theta)$, hence, $\fth(x)$ is strictly increasing.
    \item The function $\fth'(x)$ is negative in $(\theta, 1)$, hence, $\fth(x)$ is strictly decreasing.
\end{itemize}
\end{lemma}
\begin{proof}
Let $\bfth$ be a composite symbol in $\compsinbase$. We have that
$$
\fth(x)=\mathbf{P}[\Theta(\mathbf{X}^{\boldsymbol{x}}_n)=\bfth]
=\binom{n}{{{\theta}}n}{{x}}^{{{\theta}}n}(1-{{x}})^{(1-{{\theta}})n}.$$
The function $f$ is differentiable in $(0,1)$, and therefore,
\begin{align*}
\fth'(x)=\binom{n}{{{\theta}}n}
[&{{\theta}}n{{x}}^{{{\theta}}n-1}(1-{{x}})^{(1-{{\theta}})n}-(1-{{\theta}})n{{x}}^{{{\theta}}n}(1-{{x}})^{(1-{{\theta}})n-1}].
\end{align*}
Hence,
\begin{align*}
\fth'(x)=&\binom{n}{{{\theta}}n}
(n{{x}}^{{{\theta}}n-1}(1-{{x}})^{(1-{{\theta}})n-1})({\theta}(1-x)-(1-\theta)x)
=\binom{n}{{{\theta}}n}
(n{{x}}^{{{\theta}}n-1}(1-{{x}})^{(1-{{\theta}})n-1})({\theta}-x).
\end{align*}
Since
\[
\binom{n}{{{\theta}}n}
(n{{x}}^{{{\theta}}n-1}(1-{{x}})^{(1-{{\theta}})n-1})>0,
\]
we have that for $x>\theta$
\[
\fth'(x)<0\Rightarrow f \mbox{ strictly decreases.}
\]
For $x<\theta$
\[
\fth'(x)>0\Rightarrow f \mbox{ strictly increases.}
\]
\end{proof}
\begin{lemma}\label{lmadregnei}
Let $\mathcal{C}$ be an $(m,2)$-composite code $\mathcal{C}=\{\bfc_1,\dots,\bfc_m\}$. For every $n\in\N$ and for every $\bfth\in\Omega_n^2$
\begin{enumerate}
    \item If $\theta\in[c_i,c_{i+1}]$ then 
\[
\Dmld(\bfth)\in\{\bfc_i,\bfc_{i+1}\}.
\]
\item If $\theta\in[0,c_1]$ then 
\[
\Dmld(\bfth)=\bfc_1.
\]
\item If $\theta\in[c_m,1]$ then 
\[
\Dmld(\bfth)=\bfc_m.
\]
\end{enumerate}

\end{lemma}
\begin{proof}
$(1)$

    Let $\bfth\in\Omega_n^2$ such that $\theta\in[c_i,c_{i+1}]$. If $\bfth\in\{\bfc_i,\bfc_{i+1}\}$ using \autoref{cor:max_in_self} it holds that
    \[
    \Dmld(\bfth)\in\{\bfc_i,\bfc_{i+1}\}.
    \]
    Otherwise, it holds that $\theta\in(c_i,c_{i+1})$. Using Lemma \autoref{lmaincdec} for every $j>i+1$ it holds that
    \[
    \fth(\bfc_j)<\fth(\bfc_{i+1}),
    \]
    hence,
    \[
    \mathbf{P}[\Theta(\mathbf{X}^{\bfc_{i+1}}_n)=\bfth]>\mathbf{P}[\Theta(\mathbf{X}^{\bfc_j}_n)=\bfth].
    \]
    For every $j<i$ it holds that
    \[
    \fth(\bfc_j)<\fth(\bfc_{i}),
    \]
    hence,
    \[
    \mathbf{P}[\Theta(\mathbf{X}^{\bfc_{j}}_n)=\bfth]<\mathbf{P}[\Theta(\mathbf{X}^{\bfc_i}_n)=\bfth].
    \]
    Therefore, for every $n$ it holds that
    \[
    \Dmld(\bfth, n)= \mbox{argmax}_{\boldsymbol{c}\in\mathcal{C}}{\mathbf{P}[\Theta(\mathbf{X}^{\boldsymbol{c}}_n)=\bfth]}\in\{\bfc_i,\bfc_{i+1}\}.
    \]

    $(2)$

    Let $\bfth\in\Omega_n^2$ such that $\theta\in[0,c_1]$. If $\bfth=\bfc_1$ using \autoref{cor:max_in_self} it holds that
    \[
    \Dmld(\bfth)=\bfc_1.
    \]
    If $c_1=0$ then $\bfth=\bfc_1$, hence, $c_1\neq0$ and it holds that $\theta\in[0,c_1)$. If $\theta=0$ for every $i$ it holds that
    \[
    \mathbf{P}[\Theta(\mathbf{X}^{\bfc_{i}}_n)=\bfth]=(1-c_i)^n.
    \]
    Since for every $i>1$ it holds that $c_1<c_i$ we have that
    \[\mbox{argmax}_{\boldsymbol{c}\in\mathcal{C}}{\mathbf{P}[\Theta(\mathbf{X}^{\boldsymbol{c}}_n)=\bfth]}=\bfc_1.
    \]
    If $\theta\neq0$ using Lemma \autoref{lmaincdec} for every $j>1$ holds that
    \[
    \fth(\bfc_j)<\fth(\bfc_1),
    \]
    hence,
    \[
    \mathbf{P}[\Theta(\mathbf{X}^{\bfc_{1}}_n)=\bfth]>\mathbf{P}[\Theta(\mathbf{X}^{\bfc_j}_n)=\bfth].
    \]
    Therefore, for every $n$ it holds that
    \[
    \Dmld(\bfth, n)= \mbox{argmax}_{\boldsymbol{c}\in\mathcal{C}}{\mathbf{P}[\Theta(\mathbf{X}^{\boldsymbol{c}}_n)=\bfth]}=\bfc_1.
    \]

    $(3)$

    Let $\bfth\in\Omega_n^2$ such that $\theta\in[c_m,1]$. If $\bfth=\bfc_m$ using \autoref{cor:max_in_self} it holds that
    \[
    \Dmld(\bfth)=\bfc_m.
    \]
    If $c_m=1$ then $\bfth=\bfc_m$, hence, $c_m\neq1$ and it holds that $\theta\in[c_m,1)$.
    If $\theta=1$ for every $i$ it holds that
    \[
    \mathbf{P}[\Theta(\mathbf{X}^{\bfc_{i}}_n)=\bfth]=c_i^n.
    \]
    Since for every $i<m$ it holds that $c_i<c_m$ we have that
    \[\mbox{argmax}_{\boldsymbol{c}\in\mathcal{C}}{\mathbf{P}[\Theta(\mathbf{X}^{\boldsymbol{c}}_n)=\bfth]}=\bfc_m.
    \]
    If $\theta\neq1$ using Lemma \autoref{lmaincdec} for every $j<m$ holds that
    \[
    \fth(\bfc_j)<\fth(\bfc_m),
    \]
    hence,
    \[
    \mathbf{P}[\Theta(\mathbf{X}^{\bfc_{m}}_n)=\bfth]>\mathbf{P}[\Theta(\mathbf{X}^{\bfc_j}_n)=\bfth].
    \]
    Therefore, for every $n$ it holds that
    \[
    \Dmld(\bfth, n)= \mbox{argmax}_{\boldsymbol{c}\in\mathcal{C}}{\mathbf{P}[\Theta(\mathbf{X}^{\boldsymbol{c}}_n)=\bfth]}=\bfc_m.
    \]
    \end{proof}
Now we have that for any $(m,2)$-composite code $\mathcal{C}=\{\bfc_1,\dots,\bfc_m\}$ the DR of $\bfc_i\in\cC$ is affected only by its neighbours, i.e., $\bfc_{i-1}, \bfc_{i+1}$ if exist.
By placing the code $\cC$ as points on the line $[0,1]$, such that $\bfc_i$ is on $c_i$, any $\bfth\in\Omega_n^2$ can be decoded only to the neighbours from the right and from the left of $\theta$ on the line.

For every $x,y\in(0,1)$ denote 
\[
\xi(x,y)=\frac{\log(\frac{x(1-y)}{(1-x)y})}{\log(\frac{x}{y})}.
\]

\begin{lemma}\label{lmathold}
Let $\bfx,\bfy\in\compbin\setminus{E_2}$ such that $x<y$, for every $\bfth\in\ombinnsinbase$ it holds that
\[
\mathbf{P}[\Theta(\mathbf{X}^{\bfx}_n)=\bfth]>
 \mathbf{P}[\Theta(\mathbf{X}^{\bfy}_n)=\bfth] \mbox{ iff }
 \bfth<\xi({x,y}).
\]
Furthermore,
\[
\mathbf{P}[\Theta(\mathbf{X}^{\bfx}_n)=\bfth]<
 \mathbf{P}[\Theta(\mathbf{X}^{\bfy}_n)=\bfth] \mbox{ iff }
 \bfth>\xi({x,y}).
\]
\end{lemma}
\begin{proof}
It holds that
\begin{align*}
\mathbf{P}[\Theta(\mathbf{X}^{\bfx}_n)=\bfth]=
\binom{n}{{\theta}n}{x}^{{\theta}n}(1-{x})^{(1-{\theta})n},
\mathbf{P}[\Theta(\mathbf{X}^{\bfy}_n)=\bfth]=
\binom{n}{{\theta}n}{y}^{{\theta}n}(1-{y})^{(1-{\theta})n}.
\end{align*}
And now, 
\begin{gather*}
\mathbf{P}[\Theta(\mathbf{X}^{\bfx}_n)=\bfth]>
 \mathbf{P}[\Theta(\mathbf{X}^{\bfy}_n)=\bfth], \\
 \mbox{ if and only if}\\
\binom{n}{{\theta}n}{x}^{{\theta}n}(1-{x})^{(1-{\theta})n}>
\binom{n}{{\theta}n}{y}^{{\theta}n}(1-{y})^{(1-{\theta})n}, \\
 \mbox{ if and only if}\\
{x}^{{\theta}n}(1-{x})^{(1-{\theta})n}>
{y}^{{\theta}n}(1-{y})^{(1-{\theta})n}, \\
 \mbox{ if and only if}\\
(\frac{x(1-y)}{(1-x)y})^{\theta}>\frac{1-y}{1-x}.
\end{gather*}
Since $x, y, \theta \in(0,1)$ all the steps above are allowed. Furthermore we have that
\[
\frac{1-y}{1-x}<1,\frac{x}{y}<1.
\]
Finally we get
\begin{gather*}
\mathbf{P}[\Theta(\mathbf{X}^{x}_n)=\bfth]>
 \mathbf{P}[\Theta(\mathbf{X}^{y}_n)=\bfth],\\
 \mbox{ if and only if}\\
 \theta<\frac{\log(\frac{x}{y})}{\log((\frac{x(1-y)}{(1-x)y})))}=\xi({x,y}).
\end{gather*}
An identical proof with the opposite inequality results
\begin{gather*}
\mathbf{P}[\Theta(\mathbf{X}^{x}_n)=\bfth]<
 \mathbf{P}[\Theta(\mathbf{X}^{y}_n)=\bfth],\\
 \mbox{ if and only if}\\
 \theta>\frac{\log(\frac{x}{y})}{\log((\frac{x(1-y)}{(1-x)y})))}=\xi({x,y}).
\end{gather*}
\end{proof}

\begin{lemma}
Let $x,y\in(0,1)$ such that $x<y$, for every $y\leq y'<1$ it holds that
\[
\xi({x,y})<\xi({x,y'})
\]
\end{lemma}
\begin{proof}
From Lemma \autoref{lmathold} we have that for every $\bfth\in\ombinnsinbase$ it holds that
\begin{align*}
&\mathbf{P}[\Theta(\mathbf{X}^{\bfx}_n)=\bfth]>
 \mathbf{P}[\Theta(\mathbf{X}^{\bfy}_n)=\bfth] \mbox{ iff }
 \bfth<\xi({x,y}),\\
 &\mathbf{P}[\Theta(\mathbf{X}^{\bfx}_n)=\bfth]>
 \mathbf{P}[\Theta(\mathbf{X}^{\bfy'}_n)=\bfth] \mbox{ iff }
 \bfth<\xi({x,y'}).
\end{align*}
Therefore from the density of $\ombinall$ it is sufficient to prove that for every $\bfth\in\ombinnsinbase$, if $\mathbf{P}[\Theta(\mathbf{X}^{\bfx}_n)=\bfth]>
 \mathbf{P}[\Theta(\mathbf{X}^{\bfy}_n)=\bfth]$ then $\mathbf{P}[\Theta(\mathbf{X}^{\bfx}_n)=\bfth]>
 \mathbf{P}[\Theta(\mathbf{X}^{\bfy'}_n)=\bfth]$.
 Let $\bfth\in\ombinnsinbase$, if $\theta\in(0,y)$ using Lemma \autoref{lmaincdec} we have that since $y<y'$ it holds that
$ \mathbf{P}[\Theta(\mathbf{X}^{\bfy'}_n)=\bfth]<\mathbf{P}[\Theta(\mathbf{X}^{\bfy}_n)=\bfth].$
Therefore, if $\mathbf{P}[\Theta(\mathbf{X}^{\bfx}_n)=\bfth]>
 \mathbf{P}[\Theta(\mathbf{X}^{\bfy}_n)=\bfth]$ then it holds that 
 \begin{align*}
\mathbf{P}[\Theta(\mathbf{X}^{\bfx}_n)=\bfth]>
 \mathbf{P}[\Theta(\mathbf{X}^{\bfy}_n)=\bfth]
 >
 \mathbf{P}[\Theta(\mathbf{X}^{\bfy'}_n)=\bfth].
 \end{align*}
 If $\theta\in(y,1)$ using Lemma \autoref{lmaincdec} we have that since $x<y$ it holds that
 \[
 \mathbf{P}[\Theta(\mathbf{X}^{\bfx}_n)=\bfth]<\mathbf{P}[\Theta(\mathbf{X}^{\bfy}_n)=\bfth].
 \]
\end{proof}
We denote by $\bfchi^{\bfx}_{\bfy}$ all the composite symbols $\bfrh\in\compbin$ such that it is more likely that $\bfx$ has generated them rather than $\bfy$, i.e.,
\[
\mathbf{P}[\Theta(\mathbf{X}^{\bfx}_n)=\bfrh]>\mathbf{P}[\Theta(\mathbf{X}^{\bfy}_n)=\bfrh].
\]
Furthermore, we denote by $\chi^{\bfx}_{\bfy}$ the values of the composite symbols in $\bfchi^{\bfx}_{\bfy}$.
It is important to note that since the inequality does not depend on $n$, it is enough to have a specific $n$ with which the inequality holds. 

\begin{lemma}
For every $x,y\in[0,1]$ it holds that
\[
\chi^{\bfx}_{\bfy}=\begin{cases}
    [0,\xi({x,y})),& \text{if } x<y \mbox{ and }x,y\notin\{0,1\}\\
    \{0\}& \text{if } x<y \mbox{ and } x=0\\
    [0,1),& \text{if } x<y \mbox{ and } x\neq0, y=1\\
    (\xi({x,y}),1],& \text{if } x>y \mbox{ and }x,y\notin\{0,1\}\\
    \{1\}& \text{if } x>y \mbox{ and } x=1\\
    (0,1],& \text{if } x>y \mbox{ and } x\neq1, y=0\\
    \phi,& \text{if } x=y \\
    \end{cases}.
\]
\end{lemma}
\begin{proof}
    Let $x,y\in[0,1]$, 
    
$(7)$
if $x=y$. For every $\bfrh\in\compbin$ it holds that
    \[
\mathbf{P}[\Theta(\mathbf{X}^{\bfx}_n)=\bfrh]=\mathbf{P}[\Theta(\mathbf{X}^{\bfy}_n)=\bfrh],
\]
therefore, $\chi^{\bfx}_{\bfy}=\phi$. 
    
$(2)$
If $x<y$ and $x=0$. It holds that
\[
\mathbf{P}[\Theta(\mathbf{X}^{\bfx}_n)=\boldsymbol{0}]=1.
\]
For every $\bfrh\in\compbin\setminus\{\boldsymbol{0}\}$ it holds that
\[
\mathbf{P}[\Theta(\mathbf{X}^{\bfx}_n)=\bfrh]=0.
\]
Since
\[
\mathbf{P}[\Theta(\mathbf{X}^{\bfy}_n)=\boldsymbol{0}]=(1-y)^n<1,
\]
we have that $\chi^{\bfx}_{\bfy}=\{0\}$.

$(3)$
If $x<y, x\neq0, y=1$. It holds that
\[
\mathbf{P}[\Theta(\mathbf{X}^{\bfy}_n)=\boldsymbol{1}]=1.
\]
For every $\bfrh\in\compbin\setminus\{\boldsymbol{1}\}$ it holds that
\[
\mathbf{P}[\Theta(\mathbf{X}^{\bfy}_n)=\bfrh]=0,
\]
while
\[
\mathbf{P}[\Theta(\mathbf{X}^{\bfx}_n)=\bfrh]>0,
\]
hence, $\chi^{\bfx}_{\bfy}=[0,1)$.

$(1)$
If $x<y, x,y\notin\{0,1\}$. For $\bfx$ it holds that,
\begin{align*}
    \mathbf{P}[\Theta(\mathbf{X}^{\bfx}_n)=\boldsymbol{0}]=1-x,
    \mathbf{P}[\Theta(\mathbf{X}^{\bfx}_n)=\boldsymbol{1}]=x.
\end{align*}
For $\bfy$ it holds that,
\begin{align*}
    \mathbf{P}[\Theta(\mathbf{X}^{\bfy}_n)=\boldsymbol{0}]=1-y,
    \mathbf{P}[\Theta(\mathbf{X}^{\bfy}_n)=\boldsymbol{1}]=y.
\end{align*}
We assume that $x<y$, therefore $1-y<1-x$, thus, we get that $0\in\chi^{\bfx}_{\bfy}$ and $1\notin\chi^{\bfx}_{\bfy}$.
According to Lemma \autoref{lmathold} for every $\bfth\in\ombinnsinbase$ it holds that
\[
\mathbf{P}[\Theta(\mathbf{X}^{\bfx}_n)=\bfth]>
 \mathbf{P}[\Theta(\mathbf{X}^{\bfy}_n)=\bfth] \mbox{ iff }
 \theta<\xi({x,y}).
\]
Hence,
\[
\chi^{\bfx}_{\bfy}=\{0\}\cup(0,\xi({x,y}))=[0,\xi({x,y}))(1).
\]

$(5)$
If $x>y$ and $x=1$. It holds that
\[
\mathbf{P}[\Theta(\mathbf{X}^{\bfx}_n)=\boldsymbol{1}]=1.
\]
For every $\bfrh\in\compbin\setminus\{\boldsymbol{1}\}$ it holds that
\[
\mathbf{P}[\Theta(\mathbf{X}^{\bfx}_n)=\bfrh]=0.
\]
Since
\[
\mathbf{P}[\Theta(\mathbf{X}^{\bfy}_n)=\boldsymbol{1}]=y^n<1,
\]
we have that $\chi^{\bfx}_{\bfy}=\{1\}$.

$(6)$
If $x>y, x\neq1, y=0$. It holds that
\[
\mathbf{P}[\Theta(\mathbf{X}^{\bfy}_n)=\boldsymbol{0}]=1.
\]
For every $\bfrh\in\compbin\setminus\{\boldsymbol{0}\}$ it holds that
\[
\mathbf{P}[\Theta(\mathbf{X}^{\bfy}_n)=\bfrh]=0.
\]
while
\[
\mathbf{P}[\Theta(\mathbf{X}^{\bfx}_n)=\bfrh]>0,
\]
hence, $\chi^{\bfx}_{\bfy}=(0,1]$.

$(4)$
If $x>y, x,y\notin\{0,1\}$. For $\bfx$ it holds that,
\begin{align*}
    \mathbf{P}[\Theta(\mathbf{X}^{\bfx}_n)=\boldsymbol{0}]=1-x,
    \mathbf{P}[\Theta(\mathbf{X}^{\bfx}_n)=\boldsymbol{1}]=x.
\end{align*}
For $\bfy$ it holds that,
\begin{align*}
    \mathbf{P}[\Theta(\mathbf{X}^{\bfy}_n)=\boldsymbol{0}]=1-y,
    \mathbf{P}[\Theta(\mathbf{X}^{\bfy}_n)=\boldsymbol{1}]=y.
\end{align*}
We assume that $x>y$, therefore $1-y>1-x$, thus, we get that $1\in\chi^{\bfx}_{\bfy}$ and $0\notin\chi^{\bfx}_{\bfy}$.
According to Lemma \autoref{lmathold} for $x>y$ it holds that for every $\bfth\in\ombinnsinbase$ it holds that
\[
\mathbf{P}[\Theta(\mathbf{X}^{\bfx}_n)=\bfth]>
 \mathbf{P}[\Theta(\mathbf{X}^{\bfy}_n)=\bfth] \mbox{ iff }
 \theta>\xi({x,y}).
\]
Hence,
\[
\chi^{\bfx}_{\bfy}=(\xi({x,y}), 1)\bigcup\{1\}=(\xi({x,y}),1].
\]
\end{proof}

We expand the definition of $\bfchi^{\bfx}_{\bfy}$ to subsets $A\subseteq\compbin$ by
\[
\bfchi^{\bfx}_{Y}=\bigcap_{\bfy\in Y}{\bfchi^{\bfx}_{\bfy}}.
\]

\begin{lemma}\label{lmaneig}
    Let $\mathcal{C}$ be an $(m,2)$-composite code $\mathcal{C}=\{\bfc_1,\dots,\bfc_m\}$ for $m\geq2$. It holds that
\begin{itemize}
    \item For every $1<i<m$ 
    \[\drmld(\bfc_i)=\ombinall\bigcap\bfchi^{\bfc_i}_{\{\bfc_{i+1},\bfc_{i-1}\}}.
    \] 
    \item For $i=1$
    \[\drmld(\bfc_1)=\ombinall\bigcap\bfchi^{\bfc_1}_{\bfc_2}.
    \] 
    \item For $i=m$
    \[\drmld(\bfc_m)=\ombinall\bigcap\bfchi^{\bfc_m}_{\bfc_{m-1}}.
    \] 
\end{itemize}
\end{lemma}

\begin{lemma}\label{lmaexpandright}
    Let $\bfx,\bfy\in\compsinbase$ be such that $x<y$. Then for every $y\leq y'$ it holds that
    \[    \bfchi^{\bfx}_{\bfy}\subseteq\bfchi^{\bfx}_{\bfy'}.
    \]
\end{lemma}
\begin{proof}
    Let $\bfth\in\bfchi^{\bfx}_{\bfy}$ by definition it holds that
\[
\mathbf{P}[\Theta(\mathbf{X}^{\bfx}_n)=\bfth]>\mathbf{P}[\Theta(\mathbf{X}^{\bfy}_n)=\bfth].
\]
if $\theta\in[0,y)$ using Lemma \autoref{lmaincdec} we have that since $y<y'$ it holds that $
 \mathbf{P}[\Theta(\mathbf{X}^{\bfy'}_n)=\bfth]<\mathbf{P}[\Theta(\mathbf{X}^{\bfy}_n)=\bfth].$
 Since $\mathbf{P}[\Theta(\mathbf{X}^{\bfx}_n)=\bfth]>
 \mathbf{P}[\Theta(\mathbf{X}^{\bfy}_n)=\bfth]$ then it holds that 
 \begin{align*}
\mathbf{P}[\Theta(\mathbf{X}^{\bfx}_n)=\bfth]>
 \mathbf{P}[\Theta(\mathbf{X}^{\bfy}_n)=\bfth]
 >
 \mathbf{P}[\Theta(\mathbf{X}^{\bfy'}_n)=\bfth].
 \end{align*}
 Hence, $\bfth\in\bfchi^{\bfx}_{\bfy'}$
 Assume towards contradiction that $\theta\in(y,1]$ using Lemma  \autoref{lmaincdec} we have that since $x<y$ it holds that
 \[
 \mathbf{P}[\Theta(\mathbf{X}^{\bfx}_n)=\bfth]<\mathbf{P}[\Theta(\mathbf{X}^{\bfy}_n)=\bfth],
 \]
 despite that $\bfth\in\bfchi^{\bfx}_{\bfy}$, and hence we get a contradiction.
\end{proof}

\begin{lemma}\label{lmaadjprob}
    For every $A\subseteq\compbin$ and $\bfrh\in\compbin$ it holds that
    $$\mathbf{P}[\Theta(\mathbf{X}^{\bfrh}_n)\in A]=\mathbf{P}[\Theta(\mathbf{X}^{\overline{\bfrh}}_n)\in \overline{A}].
    $$
\end{lemma}
\begin{proof}
    It holds that
    \begin{align*}
    \mathbf{P}[\Theta(\mathbf{X}^{\bfrh}_n)\in A]=\sum_{\bfa\in A}\mathbf{P}[\Theta(\mathbf{X}^{\bfrh}_n)= \bfa]
    =\sum_{\bfa\in A}\binom{n}{an}\rho^{an}(1-\rho)^{(1-a)n}.
    \end{align*}
    Moreover, it holds that
    \begin{align*}
    \sum_{\bfa\in A}\binom{n}{an}\rho^{an}(1-\rho)^{(1-a)n}=\sum_{\bfa\in A}\binom{n}{(1-a)n}(1-\rho)^{(1-a)n}\rho^{an}
    =\sum_{\bfa\in A}\mathbf{P}[\Theta(\mathbf{X}^{\overline{\bfrh}}_n)= \overline{\bfa}]
    =\mathbf{P}[\Theta(\mathbf{X}^{\overline{\bfrh}}_n)\in \overline{A}].
    \end{align*}
\end{proof}

\begin{lemma}\label{lmaadjset}
    For every $\bfrh\in\compbin$ and $A\subseteq\compbin$ it holds that
    \[    \bfchi^{\bfrh}_{A}=\overline{\bfchi^{\overline{\bfrh}}_{\overline{A}}}.
    \]
\end{lemma}
\begin{proof}
It is sufficient to prove that $\bfchi^{\bfrh}_{A}\subseteq\overline{\bfchi^{\overline{\bfrh}}_{\overline{A}}}$, since $\overline{\overline{\bfchi^{\overline{\overline{\bfrh}}}_{\overline{\overline{A}}}}}=\bfchi^{\bfrh}_{A}.$
Let $\bfth\in\bfchi^{\bfrh}_{A}$, by definition for every $\bfa\in A$ it holds that
 \[
 \mathbf{P}[\Theta(\mathbf{X}^{\bfrh}_n)=\bfth]>\mathbf{P}[\Theta(\mathbf{X}^{\bfa}_n)=\bfth].
 \]
From Lemma \autoref{lmaadjprob} we have that
\begin{align*}
 \mathbf{P}[\Theta(\mathbf{X}^{\bfrh}_n)=\bfth]= \mathbf{P}[\Theta(\mathbf{X}^{\overline{\bfrh}}_n)=\overline{\bfth}],
 \mathbf{P}[\Theta(\mathbf{X}^{\bfa}_n)=\bfth]= \mathbf{P}[\Theta(\mathbf{X}^{\overline{\bfa}}_n)=\overline{\bfth}].
 \end{align*}
 Therefore, for every $\bfa\in A$ it holds that
 \[
 \mathbf{P}[\Theta(\mathbf{X}^{\overline{\bfrh}}_n)=\overline{\bfth}]>\mathbf{P}[\Theta(\mathbf{X}^{\overline{\bfa}}_n)=\overline{\bfth}],
 \]
 hence, $\overline{\bfth}\in{\bfchi^{\overline{\bfrh}}_{\overline{A}}}$, therefore, ${\bfth}\in\overline{{\bfchi^{\overline{\bfrh}}_{\overline{A}}}}$.
\end{proof}
Consequently, if $A=\{\bfrh'\}$ it holds that
    \[  \bfchi^{\bfrh}_{\bfrh'}=\overline{\bfchi^{\overline{\bfrh}}_{\overline{\bfrh'}}}.
    \]

\begin{lemma}\label{lmaadjDR}
Let $\mathcal{C}$ be an $(m,2)$-composite code $\mathcal{C}=\{\bfc_1,\dots,\bfc_m\}$, it holds for any $1\leq i\leq m$ that
\[
\drmldadj(\overline{\bfc_i})=\overline{\drmld(\bfc_i)}
\]
\end{lemma}
\begin{proof}
One can observe that for any $\bfc\in\cC$ it holds that
\[
\drmld(\bfc_i)=\bfchi^{\bfc}_{\cC\setminus\{\bfc\}},\drmldadj(\overline{\bfc_i})=\bfchi^{\overline{\bfc}}_{\overline{\cC}\setminus\{\overline{\bfc}\}}
\]
Using Lemma \autoref{lmaadjset} we have that for every $\bfc_i\in\cC$ it holds that
\begin{align*}
    \drmldadj(\overline{\bfc_i})=\bfchi^{\overline{\bfc}}_{\overline{\cC}\setminus\{\overline{\bfc}\}}
    =\overline{\bfchi^{\bfc}_{\cC\setminus\{\bfc\}}}=\overline{\drmld(\bfc_i)}.
\end{align*}
\end{proof}

\begin{lemma}\label{lmaadjsame}
Let $\mathcal{C}$ be an $(m,2)$-composite code $\mathcal{C}=\{\bfc_1,\dots,\bfc_m\}$, it holds for any $n\in\N$ that
\[
\fmin(\overline{\cC},n,\Dmldadj)=\fmin(\cC,n,\Dmld).
\]
\end{lemma}
\begin{proof}
    Using Lemma \autoref{lmaadjDR} we have that for every $\bfc_i\in\cC$ it holds that
    \[
    \drmldadj(\overline{\bfc_i})=\overline{\drmld(\bfc_i)}.
    \]
    It is known that
    \begin{align*}
        &\psucc(\cC,n,\Dmld,\bfc_{i})=\mathbf{P}[(\Theta(\mathbf{X}^{\bfc_i}_{n}))\in\drmld(\bfc_i)],\psucc(\overline{\cC},n,\Dmldadj,\overline{\bfc_{i}})=\mathbf{P}[(\Theta(\mathbf{X}^{\overline{\bfc_i)}}_{n}))\in\drmldadj(\overline{\bfc_i)}].
    \end{align*}
    Using Lemma \autoref{lmaadjprob} we have that for every $\bfc_i\in\cC$ it holds that
    \begin{align*}
        \psucc(\cC,n,\Dmld,\bfc_{i})=&\mathbf{P}[\Theta(\mathbf{X}^{\bfc_i}_{n})\in\drmld(\bfc_i)]        =\mathbf{P}[\Theta(\mathbf{X}^{\overline{\bfc_i}}_{n})\in\overline{\drmld(\bfc_i)}]
        =\mathbf{P}[(\Theta(\mathbf{X}^{\overline{\bfc_i)}}_{n}))\in\drmldadj(\overline{\bfc_i)}]
        =\psucc(\overline{\cC},n,\Dmldadj,\overline{\bfc_{i}}).
    \end{align*}
    Now we have that
    \begin{align*}
    \fmin(\mathcal{C},n,\Dmld)=\min_{\boldsymbol{c}\in\mathcal{C}}\psucc(\mathcal{C},n,\Dmld,\boldsymbol{c})
    =\min_{\overline{\boldsymbol{c}}\in\overline{\cC}}\psucc(\overline{\cC},n,\Dmldadj,\overline{\boldsymbol{c}})
    =\fmin(\overline{\cC},n,\Dmldadj).
    \end{align*}
\end{proof}

\begin{lemma}\label{lmasymequ}
Let $\mathcal{C}\in\symcode$ be a symmetric $(m,2)$-composite code $\cC=\{\bfc_1,\dots,\bfc_m\}$, for every $1\leq i\leq m$ it holds that $\overline{\bfc_i}=\bfc_{m+1-i}$, furthermore,
\begin{align*}
\psucc(\cC,n,\Dmld,\bfc_{i})=&\psucc(\cC,n,\Dmld,\overline{\bfc_{i}})
=\psucc(\cC,n,\Dmld,\bfc_{m+1-i}).
\end{align*}
\end{lemma}
\begin{proof}
Let $\mathcal{C}\in\symcode$ be a symmetric $(m,2)$-composite code $\cC=\{\bfc_1,\dots,\bfc_m\}$.
    First we will prove that for every $1\leq i\leq m$ it holds that $\overline{\bfc_i}=\bfc_{m+1-i}$.
    It is known that
    \[
    c_1<\dots<c_m.
    \]
    From the definition of symmetric symbol we have that
    \[
    \overline{c_m}>\dots>\overline{c_1}.
    \]
    Since $\cC=\overline{\cC}$ we have that for every $1\leq i\leq m$ it holds that $\overline{\bfc_i}=\bfc_{m+1-i}$.
    For every $1\leq i\leq m$ it holds that
    \[
    \psucc(\cC',n,\Dmld,\bfc_{i})=\mathbf{P}[\Theta(\mathbf{X}^{\bfc_i}_n)\in \drmld(\bfc_i)].
    \]
    Using Lemma \autoref{lmaadjprob} it is sufficient to prove that for every $1\leq i\leq m$ it holds that 
    $$\overline{\drmld(\bfc_{n-i})}=\drmld(\bfc_{i}).$$
    For every $1<i<m$ it holds from Lemma \autoref{lmaneig} that
    \begin{align*}
    \drmld(\bfc_i)=&\ombinall\bigcap\bfchi^{\bfc_i}_{\{\bfc_{i+1},\bfc_{i-1}\}},
\drmld(\bfc_{m+1-i})=\ombinall\bigcap\bfchi^{\bfc_{m+1-i}}_{\{\bfc_{m-i},\bfc_{m+2-i}\}}.
    \end{align*}
    Therefore,
    $$\overline{\drmld(\bfc_{m+1-i})}=\overline{\ombinall}\bigcap\overline{\bfchi^{\bfc_{m+1-i}}_{\{\bfc_{m-i},\bfc_{m+2-i}\}}}.$$
    By using Lemma \autoref{lmaadjset} we get that
    $$\overline{\drmld(\bfc_{m+1-i})}=\ombinall\bigcap\bfchi^{\bfc_{m+1-i}}_{\{\bfc_{m-i},\bfc_{m+2-i}\}}.$$
    Hence,
     $$\overline{\drmld(\bfc_{m+1-i})}=\drmld(\bfc_{i}).$$
     It holds from Lemma \autoref{lmaneig} that
     \begin{align*}
     \drmld(\bfc_1)=&\ombinall\bigcap\bfchi^{\bfc_1}_{\bfc_2},\\
    \drmld(\bfc_{m})=&\ombinall\bigcap\bfchi^{\bfc_{m}}_{\bfc_{m-1}}.
     \end{align*}
    Therefore,    $$\overline{\drmld(\bfc_{m})}=\overline{\ombinall}\bigcap\overline{\bfchi^{\bfc_{m}}_{\bfc_{m-1}}}.$$
    By using Lemma \autoref{lmaadjset} we get that    $$\overline{\drmld(\bfc_{m})}={\ombinall}\bigcap{\bfchi^{\bfc_{1}}_{\bfc_{2}}}$$
    Hence,
     $$\overline{\drmld(\bfc_{m})}=\drmld(\bfc_{1}).$$
    Furthermore,
    $$\drmld(\bfc_{m})=\overline{\overline{\drmld(\bfc_{m})}}=\overline{\drmld(\bfc_{1})}.$$
     
\end{proof}

\begin{lemma}\label{lmamakesymbetter}
Let $\mathcal{C}$ be a $(2m,2)$-composite code $\cC=\{\bfc_1,\dots,\bfc_m,\bfc_{m+1},\dots,\bfc_{2m}\}$ such that $c_m<\frac{1}{2}$ and $\overline{c_m}\geq c_{m+1}$. Denote $\cC'=\{\bfc_1,\dots,\bfc_m\}\cup\overline{\{\bfc_1,\dots,\bfc_m\}}$, i,e,
\[
\cC'=\{\bfc_1,\dots,\bfc_m,\overline{\bfc_m},\dots,\overline{\bfc_1}\}.
\]
Denote by $\Dmld,\Dmld'$ $\drmld,\drmld'$ the MLD and the DR for each code respectively, for every $n\in\N$ it holds that
\[
\fmin(\cC',n,\Dmld')\geq\fmin(\cC,n,\Dmld).
\]
\end{lemma}
\begin{proof}
    Let $\cC,\cC'$ be such codes. From Lemma \autoref{lmaneig} we can deduce that for every $1\leq i<m$ it holds that
    \[
    \psucc(\cC',n,\Dmld',\bfc_{i})=\psucc(\cC,n,\Dmld,\bfc_{i})
    \]
    The code $\cC'$ is symmetric, therefore, using Lemma \autoref{lmasymequ} it holds that for every $1\leq i\leq m$
    \[
    \psucc(\cC',n,\Dmld',\bfc_{i})=\psucc(\cC',n,\Dmld',\bfc_{2m+1-i})
    \]
    By definition it holds that
    $\fmin(\cC,n,\Dmld)=\min_{\boldsymbol{c}\in\mathcal{C}}\psucc(\mathcal{C},n,\Dmld,\boldsymbol{c}).$
    Therefore, for every $1\leq i< m$ it holds that
    $\psucc(\cC',n,\Dmld',\overline{\bfc_{i}})=\psucc(\cC',n,\Dmld',\bfc_{i})
    =\psucc(\cC,n,\Dmld,{\bfc_{i}})\geq\fmin(\cC,n,\Dmld).$
    Hence, it is sufficient to prove that $\psucc(\cC',n,\Dmld,{\bfc_{m}})\geq\fmin(\cC,n,\Dmld).$
    It is known that $
    \psucc(\cC',n,\Dmld',\bfc_{m})=\mathbf{P}[\Theta(\mathbf{X}^{\bfc_m}_n)\in \drmld'(\bfc_m)].
    $
    From Lemma \autoref{lmaneig} it holds that
    $
    \drmld(\bfc_m)=\ombinall\bigcap\bfchi^{\bfc_m}_{\{\bfc_{m+1},\bfc_{m-1}\}},
    \drmld'(\bfc_m)=\ombinall\bigcap\bfchi^{\bfc_m}_{\{\overline{\bfc_{m}},\bfc_{m-1}\}}.
    $
    Since $\overline{c_m}\geq c_{m+1}$ it holds from Lemma \autoref{lmaexpandright} that
    $
    \bfchi^{\bfc_m}_{\bfc_{m+1}}\subseteq\bfchi^{\bfc_m}_{\overline{c_m}},
    $
    hence,
    $
    \drmld(\bfc_m)\subseteq\drmld'(\bfc_m),
    $
    thus, we get that
    $\psucc(\cC',n,\Dmld',\bfc_{m})=
    \mathbf{P}[\Theta(\mathbf{X}^{\bfc_m}_n)\in \drmld'(\bfc_m)]\geq
    \mathbf{P}[\Theta(\mathbf{X}^{\bfc_m}_n)\in \drmld(\bfc_m)]
    =\psucc(\cC,n,\Dmld,\bfc_{m})\geq\fmin(\cC,n,\Dmld).
    $
\end{proof}

\sameminwithsymLm*
\begin{proof}
    Let $\cC\in\codebinconbase^{2m}$ be a binary code of size $m$ that contains $E_2$, denote $\cC=\{\bfc_1,\dots,\bfc_m,\bfc_{m+1},\dots,\bfc_{2m}\}$.
    If $c_m\geq\frac{1}{2}$ then $\overline{c_{m+1}}<\frac{1}{2}$ and from Lemma \autoref{lmaadjsame} it holds for any $n\in\N$ that
$
\fmin(\overline{\cC},n,\Dmldadj)=\fmin(\cC,n,\Dmld).
$
Therefore we can use $\overline{\cC}=\{\bfc_1',\dots,\bfc_m',\bfc_{m+1}',\dots,\bfc_{2m}'\}$ instead of $\cC$ and then $c_{m}'=\overline{c_m}<\frac{1}{2}$.
    Hence, we can assume that $c_m<\frac{1}{2}$.
    If $\overline{c_m}<c_{m+1}$ then in $\overline{\cC}$ we have that
    $\overline{c_m'}=c_{m+1}>\overline{c_m}=c_{m+1}'$, furthermore $c_m'=\overline{c_{m+1}}<c_m<\frac{1}{2}$. In summary we can assume that $c_m<\frac{1}{2}$ and that $\overline{c_m}<c_{m+1}$. From Lemma \autoref{lmamakesymbetter} we deduce that there exists $\cC\in\symevencodeconbase$ such that
    $\fmin(\mathcal{C}',n,\Dmld)\geq\fmin(\mathcal{C},n,\Dmld).
    $ 
\end{proof}

\end{appendices}


\begin{thebibliography}{9}


\bibitem{anavy_DataStorageDNA_2019} 
 L. Anavy, I. Vaknin, O. Atar, R. Amit, and Z. Yakhini, 
``Data storage in DNA with fewer synthesis cycles using composite DNA letters," \emph{Nature Biotechnology}, vol. 37, no. 10, pp. 1229–1236, 2019.

\bibitem{CYB}
D. Bar-Lev, O. Sabary, R. Gabrys, and E. Yaakobi, “Cover your bases: How to minimize 
the sequencing coverage in DNA storage systems,” arXiv preprint arXiv:2305.05656, 2023. 


\bibitem{augmented_encoding}
Y. Choi, T. Ryu,  A. Lee, H. Choi,  H. Lee,  J. Park,  S.H., Song,  S. Kim,  H. Kim,  W. Park,  and S. Kwon, 
``High information capacity DNA-based data storage with augmented encoding characters using degenerate bases,"
\emph{Scientific Reports}, vol.9, 2019. 

\bibitem{Getal13} 
N. Goldman et al.,
``Towards practical, high-capacity, low-maintenance information storage in synthesized DNA," \emph{Nature}, vol. 494, no. 7435, pp. 77--80, 2013.

\bibitem{KYW23}
A. Kobovich, E. Yaakobi, and N. Weinberger, 
``M-DAB: An input-distribution optimization algorithm for composite DNA storage by the multinomial channel," Arxiv, Sep., 2023,  http://arxiv.org/abs/2309.17193.

\bibitem{osti_1619517}
H. Lee, R. Kalhor, N. Goela, J. Bolot, and G.M. Church,
``Terminator-free template-independent enzymatic DNA synthesis for digital information storage,"
\emph{Nature Communications}, vol. 10, no. 1, Jun., 2019.


\bibitem{PGYA24}
I. Preuss, B. Galili, Z. Yakhini and L. Anavy, "Sequencing Coverage Analysis for Combinatorial DNA-Based Storage Systems," IEEE Transactions on Molecular, Biological, and Multi-Scale Communications, 2024.

\bibitem{PRYA24}
I. Preuss, M. Rosenberg, Z. Yakhini, and L. Anavy, ``Efficient DNA-based data storage using shortmer combinatorial encoding," (Jan. 2,
2024), [Online]. Available: https://www.biorxiv.org/content/10.1101/
2021.08.01.454622v2 (visited on 01/29/2024), preprint.

\bibitem{ISIT24_1}
O. Sabary, I. Preuss, R. Gabrys, Z. Yakhini, L. Anavy, and E. Yaakobi, 
``Error-correcting codes for combinatorial DNA composite," \emph{IEEE International Symposium on Information Theory}, 2024.





\bibitem{ISIT24_2}
F. Walter, O. Sabary, A. Wachter-Zeh, and E. Yaakobi, 
``Coding for composite DNA to correct substitutions, strand losses, and deletions," submitted to \emph{IEEE International Symposium on Information Theory}, 2024.

\bibitem{ZC22}
W. Zhang, Z. Chen, and Z. Wang, 
``Limited-Magnitude Error Correction for Probability Vectors in DNA Storage,"
\emph{ IEEE International Conference on Communications (ICC)}, pp. 3460--3465, 2022. 


\bibitem{RVE}
B. Eisenberg "On the expectation of the maximum of IID geometric random variables." Statistics and Probability Letters. 78. 135-143. 10.1016/j.spl.2007.05.011. 



\bibitem{WESEL}
R. D. Wesel, “Efficient binomial channel capacity computation with an application to molecular communication,” in Proc. ITA, 2018, pp. 1–5. 

\bibitem{Amit}
A. Zrihan, E. Yaakobi, and Z. Yakhini, “Studying the Cycle Complexity of DNA Synthesis” \emph{EEE Information Theory Workshop (ITW), 2024}




\end{thebibliography}
\end{document}